\documentclass[lettersize,journal]{IEEEtran}
\usepackage[hidelinks]{hyperref}

\usepackage[table]{xcolor}
\usepackage{amsmath,amsfonts}
\usepackage{algorithmic}
\usepackage{array}
\usepackage[normalem]{ulem}
\usepackage[caption=false,font=normalsize,labelfont=sf,textfont=sf]{subfig}
\usepackage{textcomp}
\usepackage{stfloats}
\usepackage{url}
\usepackage{verbatim}
\def\BibTeX{{\rm B\kern-.05em{\sc i\kern-.025em b}\kern-.08em
    T\kern-.1667em\lower.7ex\hbox{E}\kern-.125emX}}
\usepackage{balance}

\usepackage{graphicx}
\graphicspath{{./Figs/}}
\DeclareGraphicsExtensions{.pdf,.jpeg,.png,.eps}

\usepackage{booktabs}
\usepackage{url}
\usepackage{multirow}
\usepackage{tabularx}
\usepackage{siunitx} 
\usepackage{caption}
\usepackage{subcaption}
\usepackage{subcaption,amsfonts,dcolumn}
\usepackage{xspace}

\usepackage[T1]{fontenc}
\usepackage[utf8]{inputenc} 
\usepackage{listings}

\usepackage{hyperref}
\usepackage{pbox} 
\usepackage{xurl}
\usepackage{seqsplit}
\usepackage{csquotes}
\usepackage{enumerate}
\usepackage[most]{tcolorbox}
\usepackage{float}
\usepackage[export]{adjustbox}
\usepackage{tikz}
\usepackage{array}
\usepackage{makecell}
\usepackage{booktabs}
\usepackage{tabularx, array, makecell, threeparttable} 
\newcolumntype{L}[1]{>{\raggedright\arraybackslash}m{#1}} 
\newcolumntype{R}[1]{>{\raggedleft\arraybackslash}m{#1}}  
\newcolumntype{Y}{>{\raggedright\arraybackslash}X}        

\newcommand{\citep}[1]{\cite{#1}}

\newcommand{\citet}[1]{\csuse{mapcitet#1}~\cite{#1}}

\definecolor{custom-gray}{cmyk}{0,0,0,0.7,1.00}
\definecolor{rank1}{HTML}{99FF99}
\definecolor{rank2}{HTML}{FFFF99}
\definecolor{rank3}{HTML}{FFCC99}
\definecolor{rank4}{HTML}{FF9999}
\definecolor{rank5}{HTML}{DDDDDD}  
\newtcolorbox{Summary}[2][]{
    top=0.15in,
    fonttitle=\bfseries,
    colbacktitle=custom-gray,
    colback=gray!5,
    colframe=gray!40!black,
    enhanced,
    attach boxed title to top left={xshift=1.5em,yshift=-\tcboxedtitleheight/2},
    boxed title style={size=small,colback=custom-gray},
    drop shadow={black!50!white},
    title=#2,#1}

\lstdefinelanguage{json}{
  upquote=true,
  columns=fullflexible,
  showstringspaces=false,
  breaklines=true,
  numbers=left,
  numberstyle=\tiny\color{gray},
  numbersep=6pt,
  frame=single,
  framerule=0.4pt,
  framesep=2pt,
  xleftmargin=0pt,
  xrightmargin=0pt,
  backgroundcolor=\color{black!3},
  literate=
   *{0}{{{\color{blue}0}}}{1}
    {1}{{{\color{blue}1}}}{1}
    {2}{{{\color{blue}2}}}{1}
    {3}{{{\color{blue}3}}}{1}
    {4}{{{\color{blue}4}}}{1}
    {5}{{{\color{blue}5}}}{1}
    {6}{{{\color{blue}6}}}{1}
    {7}{{{\color{blue}7}}}{1}
    {8}{{{\color{blue}8}}}{1}
    {9}{{{\color{blue}9}}}{1}
    {:}{{{\color{red}:}}}{1}
    {,}{{{\color{red},}}}{1}
    {\{}{{{\color{orange}\{}}}{1}
    {\}}{{{\color{orange}\}}}}{1}
    {[}{{{\color{orange}[}}}{1}
    {]}{{{\color{orange}]}}}{1},
}

\lstdefinestyle{jsoncompact}{
  language=json,
  basicstyle=\ttfamily\footnotesize, 
  aboveskip=3pt,
  belowskip=3pt,
  captionpos=b,
  abovecaptionskip=2pt,
  belowcaptionskip=0pt,
}

\definecolor{custom-gray}{cmyk}{0, 0, 0, 0.7, 1.00}

\begin{document}
\bstctlcite{BSTcontrol}


\title{From Human-Centric to Agentic Code Review: The Impact of Different Generations of Generative AI Technology on Review Quality}


\author{
    Suzhen Zhong,
    Shayan Noei,
    Bram Adams,
    Ying Zou
    \thanks{Suzhen Zhong, Shayan Noei, and Ying Zou are with the Department of Electrical and Computer Engineering, Queen’s University, Kingston, ON K7L 3N6, Canada. E-mail: \{suzhen.zhong, s.noei, ying.zou\}@queensu.ca.}
    \thanks{Bram Adams is with the Maintenance, Construction and Intelligence of Software Lab (MCIS), School of Computing, Queen’s University, Kingston, ON K7L 3N6, Canada. E-mail: bram.adams@queensu.ca.}
}



\maketitle

\begin{abstract}
Code review helps maintain software quality before code integration, but it also imposes a substantial workload on human reviewers. As generative artificial intelligence becomes part of software development, code review is shifting from a primarily human review process toward AI-supported review processes in which large language model~(LLM) reviewers and AI agent reviewers participate alongside human reviewers. However, we still lack empirical evidence on how this transition affects review efficiency and review quality. In this paper, we study 1.02 million reviewed pull requests from 207 GitHub projects that transition across three code review eras: human-centric review, LLM-assisted review, and agentic code review. We identify three AI reviewer adoption practices: Gradual AI Adoption, Rapid LLM Adoption, and Rapid AI Agent Adoption. We further model pull request review discussions as reviewer interaction sequences to characterize how human, LLM, and AI agent reviewers collaborate during the review process. Our results show that agent-involved collaboration patterns, especially reviews initiated by AI agents or involving multiple AI agents, are associated with faster review decisions under Gradual AI Adoption and Rapid AI Agent Adoption. However, these efficiency gains do not translate into better review quality. We also find that review activity and pull request type remain important across eras, while human-AI collaboration patterns become the strongest explanatory factor for review efficiency once LLM and AI agent reviewers participate. These findings provide empirical guidance for designing AI-supported code review processes that improve efficiency without weakening review quality.

\end{abstract}
\begin{IEEEkeywords}
Agentic Code Review, Modern Code Review
\end{IEEEkeywords}

\newcommand{\rqone}{What are the common AI adoption practices and how do they relate to review quality?} 

\newcommand{\rqtwo}{What is the impact of human-AI collaboration patterns on code review quality?}

\newcommand{\rqthree}{How do human-AI collaboration patterns relate to traditional factors impacting review quality?} 

\newcommand{\motivation}{\textbf{\textit{Motivation. }}}
\newcommand{\approach}{\textbf{\textit{Approach. }}}
\newcommand{\findings}{\textbf{\textit{Results. }}}

\section{Introduction}\label{sec:introduction}


Generative AI has been rapidly integrated into software development. For example, AI-generated code now accounts for 75\% of new code at Google~\cite{google_ai_code_75}. GitHub reports that developers pushed nearly one billion commits in 2025, while public repositories using an LLM SDK grew by 178\% year over year~\cite{github_octoverse_2025}. As AI-generated code increases the volume of code changes, code review and code integration are becoming downstream bottlenecks. Recent blog posts from Google~\cite{GoogleReviewBottleneck2026} and Amazon~\cite{AmazonReviewBottleneck2026} warn that faster code generation can overwhelm human reviewers and delivery pipelines.

With this rapid progress in generative AI, developers are increasingly adopting generative AI to write and review code and automate software engineering tasks~\cite{li2025riseaiteammatessoftware,zhong2025developerllm}. Initially~(Pre-LLM era), human reviewers manually provided review feedback, potentially assisted by bots or older machine learning~(ML) tools. Later~(LLM era), early LLM reviewers assisted developers in generating natural-language feedback from code changes~\cite{rasheed2024aipoweredcodereviewllms}. More recently~(agent era), AI agent reviewers have extended the capabilities of LLMs by autonomously retrieving project context, running development tools, and verifying findings before producing feedback~\cite{CopilotAgent2025}. 
Across these three eras, code review shifts from a human-centric process to a human-AI collaboration review process in which human reviewers, LLM reviewers, and AI agent reviewers may need to collaborate to complete the code review.

Prior work finds that frequent human reviews of small code changes support efficient review decisions~\cite{rigby2013convergent,sadowski2018modern}, while reviewer workload and patch characteristics influence review time~\cite{baysal2016investigating,thongtanunam2017review}. Other studies have examined review smells~\cite{reviewsmell2022}, showing that assigning unsuitable reviewers or reviewing pull requests with too many code changes at once can reduce review quality. Recent studies find that LLM-assisted review may increase review time~\cite{cihan2025automatedcodereview}. Thus far, existing studies provide limited evidence on how code review practices and review quality evolve as teams transition from human-centric review~(without generative AI participation) to LLM-assisted and agentic review. Therefore, it remains unclear how projects utilize AI reviewers over time, how human, LLM, and AI agent reviewers collaborate throughout the review process, and which practices and factors are associated with review quality.

To answer these questions, we analyze 1.02 million pull requests~(PR) from 207 open-source projects whose code review practices have evolved from the human-centric to the LLM-assisted and agentic review eras. 
We aim to answer the following research questions (RQs):

\textbf{RQ1. \rqone} Although LLM and AI agent reviewers are increasingly used in code review, it is unclear how projects adopt these reviewers and whether different adoption practices affect review quality. Across the 207 studied projects, we identify three AI reviewer adoption practices: (1) Gradual AI Adoption, (2) Rapid LLM Adoption, and (3) Rapid AI Agent Adoption. By comparing review efficiency and review smells across the three review eras, we find that the \textit{Rapid LLM Adoption} is significantly associated with an increase in code review smell prevalence. However, \textit{Gradual AI Adoption} and \textit{Rapid AI Agent Adoption} practices show significantly more efficient review in the agent era.

\textbf{RQ2. \rqtwo} To identify which human-AI review collaboration patterns are associated with better review quality under different AI adoption practices, we examine the interactions among human, LLM, and AI agent reviewers within individual pull requests. We find that no human-AI collaboration pattern consistently outperforms human-only review in both efficiency and quality. In the agent era, reviews initiated by AI agents or involving multiple AI agents are significantly more efficient than human-only reviews in \textit{Gradual AI Adoption} and \textit{Rapid AI Agent Adoption}. However, most collaboration patterns involving LLM or AI agent reviewers show significantly higher review quality risk than human-only review.

\textbf{RQ3. \rqthree} To better understand the factors explaining the code review quality issues identified in RQ2, we build explanatory logistic regression models to assess the important factors related to review quality. We find that once LLM and AI agent reviewers join, human-AI collaboration patterns emerge as a strong explanatory factor alongside traditional review factors, especially for Review Buddies, a smell where repeated reliance on the same reviewers may narrow review perspectives. Meanwhile, agent-involved collaboration patterns under the Gradual AI Adoption and Rapid AI Agent Adoption practices are consistently associated with larger changesets, indicating that these collaboration patterns are more frequently used to review larger code changes.

Our work makes the following contributions:

\begin{itemize}
    \item We provide a large-scale empirical study of 1.02 million pull requests from 207 open-source GitHub projects across the transitions from human-centric review to LLM-assisted and agentic review. We also release a replication package~\cite{github_ai_review_evolution} containing the longitudinal dataset to support future research on AI reviewer adoption.

    \item We characterize the adoption of LLM and AI agent reviewers in projects and quantify the association between the adoption practices and review quality, giving practitioners evidence to guide their adoption of AI reviewers.

    \item We model review quality across the three generative AI review eras, jointly considering human-AI collaboration patterns and the traditional factors~(e.g., pull request characteristics, review activity). We show that human-AI collaboration patterns and these traditional review factors both remain important for explaining review quality.

    \item We present the first large-scale quantitative studies of agentic code review, analyzing how AI reviewer adoption evolves across the Pre-LLM, LLM, and Agent eras.
    
\end{itemize}
\section{Case Study Setup}\label{sec:methodology}

This section details the case study setup. We describe our data collection and analysis approaches. The detailed approaches for each RQ are presented in the Sec.~\ref{sec:RQ1}--\ref{sec:RQ3}.


\begin{figure*}[!t]
  \centering
  \includegraphics[width=0.99\textwidth]{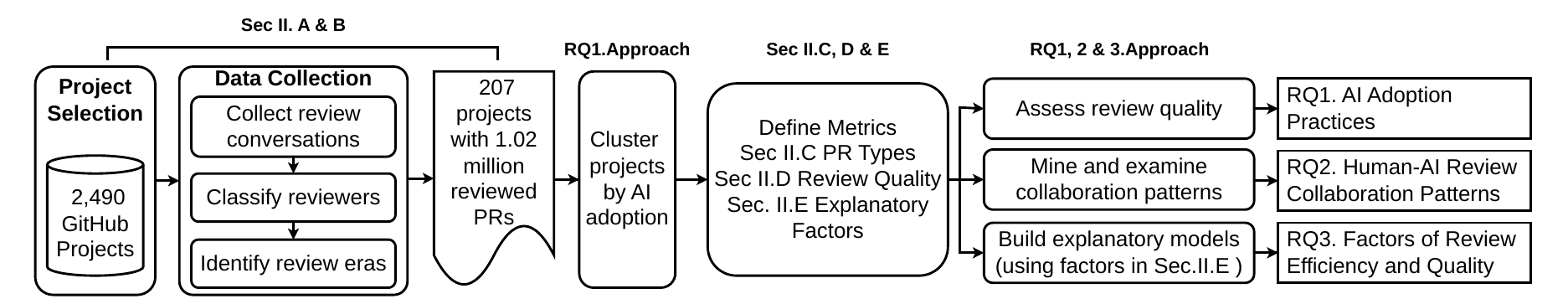}
  \caption{Overview of our approach.}
  \vspace{-10pt}
  \label{fig:RQ0_Approach}
\end{figure*}

An overview of our study is shown in Fig.~\ref{fig:RQ0_Approach}. 
Starting from 2,490 candidate GitHub projects with continuous review activity, we retain 207 with sufficient reviewed pull requests across the pre-LLM, LLM, and agent eras. We cluster the generative AI adoption series for each project into common AI reviewer adoption practices. Within each practice, we examine the pull request types~(e.g., bug fixing) that increasingly involve AI reviewers and investigate the changes in review quality across the three eras. To understand the collaboration among human, LLM, and AI agent reviewers, we identify common human-AI collaboration patterns, and compare the patterns by review efficiency and quality. Finally, we build explanatory models to assess the relationship between review quality and the human-AI collaboration patterns, alongside traditional review-process factors~(e.g., pull request characteristics).

\subsection{Project Selection}\label{sec:select_projects}

To capture code review across the transition from human-centric review to LLM-assisted and agentic review, we select active GitHub projects with continuous review activity before and after the emergence of generative AI reviewers. Using the GitHub advanced search~\cite{github_search}, we identify initial candidate projects with continuous review activity from May 2022 to February 2026. We retain projects that meet the criteria below:

\begin{itemize}
    \item at least 100 stars~\cite{unmaintain_project_github}, to ensure that a project has sufficient community adoption and is actively maintained;
    \item created before May 2022, providing at least six months of review history before the public release of ChatGPT in November 2022~\cite{wikipedia_chatgpt}; and 
    \item at least one reviewed pull request per month from May 2022 to February 2026, ensuring continuous review activity throughout the observation window.
\end{itemize}

The criteria yield 2,490 candidate projects, with continuous activity from May 2022 through February 2026.

\subsection{Data Collection and Preprocessing}

For each of the 2,490 candidate projects, we use the GitHub REST API~\cite{github_rest_api} to collect their reviewed pull requests and associated review conversations, including review events, reviewer identities, timestamps, pull request decisions after review~(i.e., accepted if merged or rejected if closed without merge), and review comments from each reviewer.

\textbf{Labeling reviewers.}
To distinguish human reviewers from automated reviewers, we label each reviewer account with the reviewer type represented by the tool behind the account at the time of review. We first use the GitHub REST API~\cite{github_rest_api} to distinguish human and bot accounts. For bot accounts, we manually inspect the official web documentation of the corresponding tool to determine whether the account represents a rule-based bot~(e.g., GitHub Actions~\cite{GithubActions}), a traditional machine-learning review tool~(e.g., Amazon CodeGuru Reviewer~\cite{AWSCodeGuru}), an LLM reviewer~(e.g., LlamaPReview~\cite{llamapreview}), or an AI agent reviewer~(e.g., Claude Code~\cite{ClaudeCode2025}). To validate that the AI agent reviewer labels reflect agentic review behavior, we manually inspected a statistically representative sample of 384 agent-era pull requests involving AI agent reviewers, corresponding to a 95\% confidence level and a 5\% margin of error~\cite{israel1992determining}. Among these sampled pull requests, 360 contained evidence of agentic behavior, such as planning, retrieval of repository context, or commit actions. For the remaining 24 sampled cases with only natural-language comments, we checked project documentation and configuration files and found no evidence that agentic review was disabled. Since official documentation identifies these reviewers as AI agents, we label these cases as involving AI agent reviewers.


After assigning reviewer labels, we define review eras for each project based on the first participation of each LLM or AI agent reviewer. For each project, its~\textbf{pre-LLM era} includes all reviewed pull requests before any generative AI reviewer participates. Its~\textbf{LLM era} starts with the first reviewed pull request involving an LLM reviewer and continues until the first reviewed pull request involving an AI agent reviewer. The project's~\textbf{agent era} starts with the first reviewed pull request involving an AI agent reviewer. Because review eras are defined separately for each project, each era's start time and end time vary across projects. After defining the review eras, we retain only projects with more than 400 reviewed pull requests in each era to reduce the influence of one-off or experimental AI use and to ensure sufficient review activity for within-project statistical comparisons~\cite{israel1992determining}. Finally, we obtain 1.02 million reviewed pull requests from 207 projects.

\begin{table}[t]
\centering
\caption{Pull request (PR) categories adopted from Li et al.~\cite{li2025riseaiteammatessoftware}.}
\vspace{-5pt}
\label{tab:pr_types}
\begin{tabularx}{\columnwidth}{l X}
\toprule
\textbf{Category} & \textbf{Description} \\
\midrule
Chore & Routine maintenance or configuration tasks. \\
Fix & Corrects bugs. \\
CI & Updates CI/CD pipelines or workflow configurations. \\
Feature & Adds or implements new functionality. \\
Performance & Tries to improve speed or efficiency. \\
Refactor & Restructures existing code without changing behavior. \\
Documentation & Updates or adds documentation. \\
Style & Adjusts formatting or naming conventions. \\
Test & Adds or modifies test cases. \\
Build & Modifies build process or dependencies. \\
Other & Unclassified or mixed-purpose changes. \\
\bottomrule
\end{tabularx}
\vspace{-15pt}
\end{table}

\subsection{Pull Request Types}
As shown in Fig. 1,
all three RQs use pull request type as a shared measure. We therefore describe here how each pull request is assigned a type.

Watanabe et al.~\cite{watanabe2025useagenticcodingempirical} proposed 11 pull request types for characterizing code-change purpose, with the full taxonomy shown in Table~\ref{tab:pr_types}.
Li et al.~\cite{li2025riseaiteammatessoftware} classify pull requests using this taxonomy by analyzing their titles and descriptions with GPT-4.1-mini. We apply the same PR-level classification process to assign each pull request a type. To validate the reliability of the classifications produced by GPT-4.1-mini, we manually label a statistically representative sample of 384 pull requests with 95\% confidence level and 5\% margin of error ~\cite{israel1992determining}. Comparing manual labels against LLM classifications yields a Cohen's $\kappa$ ~\cite{cohen1960coefficient} of 0.91, indicating almost perfect agreement between the human and GPT-4.1-mini classifications.

\subsection{Review Quality Evaluation}
\label{sec:review-quality-evaluation}

As projects shift from human-centric review toward agentic review, we examine whether these transitions are associated with changes in review efficiency and review quality risks. As shown in Fig.~\ref{fig:RQ0_Approach}, review efficiency and review smells are the shared review-quality measures used across all three RQs.

\begin{table}[t]
\centering
\caption{Code review smell taxonomy and detection rules adapted from Doğan and Tüzün~\cite{reviewsmell2022}.}
\label{tab:rq2_taxonomy_smell}
\vspace{-5pt}
\setlength{\tabcolsep}{1pt}
\renewcommand{\arraystretch}{1.08}
\begin{tabularx}{\columnwidth}{
  >{\raggedright\arraybackslash}p{1.28cm}
  >{\raggedright\arraybackslash}X
  >{\raggedright\arraybackslash}X}
\toprule
\textbf{Smell} & \textbf{Potential Side Effect} & \textbf{Detection Rule} \\
\midrule
Sleeping\newline Review &
Delayed decisions may cause context loss or block downstream work. &
Time from PR creation to final decision exceeds two days. \\
\midrule
Review\newline Buddies &
Repeated reliance on the same reviewers may narrow review perspectives. &
Same reviewer~(human, LLM, or AI agent) reviews at least 50\% of an author's PRs. \\
\midrule
Large\newline Changeset &
Large changes may reduce reviewer focus and feedback effectiveness. &
Code churn exceeds 500 changed lines of code. \\
\midrule
Ping-\newline Pong &
Excessive back-and-forth may indicate inefficient communication or rework. &
PR has more than three change-request iterations from human or AI reviewers. \\
\midrule
Missing\newline Context &
Missing context may weaken reviewer understanding and feedback usefulness. &
PR description is empty or lacks linked issue/context information. \\
\midrule
Lack of\newline Review &
Missing review may let defects or maintainability issues enter the codebase. &
No human, LLM, or AI agent reviewer other than the author participates. \\
\bottomrule
\end{tabularx}
\vspace{-15pt}
\end{table}


\subsubsection{Review Efficiency} To assess changes in user-visible review efficiency, we follow prior work~\cite{izquierdo2017using} and measure the number of days from the creation of a pull request to the final review decision. Since larger changes may require more review effort, we normalize the review duration by the size of the code change in thousands of lines of code (KLOC)~\cite{izquierdo2017using}:
\vspace{-5pt}
\begin{equation}
\text{Review Efficiency} = \frac{T_{\text{decision}} - T_{\text{creation}}}{\text{\# Thousand lines of code}}
\end{equation}
\textit{where $T_{\text{creation}}$ is the timestamp of the creation of the pull request and $T_{\text{decision}}$ is the timestamp of the final decision.}

\subsubsection{Review Smells} Review smells capture observable anti-patterns in the review process that may reduce review effectiveness and introduce quality risks~\cite{antipatterninreview2021}. For example, low review participation is associated with more post-release defects~\cite{mcintosh2016empirical}, and large code changes tend to receive less useful review feedback~\cite{bosu2015characteristics}. Although review smells were originally defined for human code review, they remain relevant in human-AI review because AI participation does not remove core review-process risks, such as missing context, repeated back-and-forth, and reliance on narrow feedback sources. We therefore use review smells as indicators of review-process quality risk in both human-only and human-AI reviews. Following Doğan and Tüzün~\cite{reviewsmell2022}, who define code review smells and detection rules from code review histories, we adapt six of their smells that can be operationalized using our pull request review conversations, then apply these rules to each reviewed pull request to identify the review smells, as shown in Table~\ref{tab:rq2_taxonomy_smell}.
\begin{table*}[!t]
\centering 

\caption{Independent explanatory variables used to model how reviewer collaboration patterns and alternative factors are jointly associated with review efficiency and review smells. The Abbrev. column shows the shortened metric names used in Table~\ref{tab:rq3_effi_smell_byera}.}
\vspace{-5pt}
\label{tab:rq3-metrics}
\begin{tabularx}{\linewidth}{l l@{\hspace{0.5\tabcolsep}}l l X}
\toprule
\textbf{Metric} & \textbf{Abbrev.} & \textbf{Category} & \textbf{Source} & \textbf{Description} \\
\midrule
Review era
& Era
& Review era
& This study
& The three review eras~(i.e., pre-LLM, LLM, and agent eras) \\
\midrule
PR type
& PR type
& PR Characteristics
& \cite{li2025riseaiteammatessoftware, watanabe2025useagenticcodingempirical}
& Purpose of the pull request, such as Feature, Fix, and Chore (Table~\ref{tab:pr_types}) \\

Initial code churn
& Churn
& PR Characteristics
& \cite{mcintosh2016empirical, thongtanunam2017review}
& Number of added and deleted lines when the pull request is opened \\

Initial files changed
& Files
& PR Characteristics
& \cite{mcintosh2016empirical, thongtanunam2017review}
& Number of files changed when the pull request is opened \\

Initial commit count 
& InitCommit
& PR Characteristics 
& \cite{bosu2015characteristics} 
& Number of commits present when the pull request is created \\
\midrule
Commit count
& NumCommit
& Review activity
& \cite{bosu2015characteristics}
& Total number of commits in the PR when the review process ends \\

Inline comments
& InlineThread
& Review activity
& \cite{zhong2026humanaisynergyagenticcode}
& Number of reviewer discussion threads on specific changed lines \\

Reviewer count
& ReviewerCount
& Review activity
& \cite{thongtanunam2017review}
& Number of unique reviewers of the pull request \\
\midrule
Author experience
& Author Exp
& Experience
& \cite{bosu2015characteristics, mcintosh2016empirical}
& Number of prior commits by the author in the same project \\

Human reviewer experience
& Human Exp
& Experience
& \cite{thongtanunam2017review, baysal2016investigating}
& Mean prior reviews by the PR's human reviewers in the same project \\

Bot reviewer experience
& Rule Exp
& Experience
& This study
& Mean prior reviews by the PR's rule-based bot in the same project \\

ML reviewer experience
& ML Exp
& Experience
& This study
& Mean prior reviews by the PR's ML reviewers in the same project \\

LLM reviewer experience
& LLM Exp
& Experience
& This study
& Mean prior reviews by the PR's LLM reviewers in the same project \\

Agent reviewer experience
& Agent Exp
& Experience
& This study
& Mean prior reviews by the PR's AI agent reviewers in the same project \\
\bottomrule
\end{tabularx}

\vspace{2pt}
\parbox{\textwidth}{\footnotesize *ML stands for traditional Machine Learning, which is not generative AI.}
\vspace{-20pt}
\end{table*}


\subsection{Explanatory Factors for Review Efficiency and Quality}
As shown in Fig.~\ref{fig:RQ0_Approach}, we show the explanatory factors used in the RQ3 analysis. Here, we introduces traditional review factors from prior code review studies, while Section~\ref{sec:RQ3} shows the modeling procedure. Prior work on code review with human reviewers shows that review quality is associated with review process factors, including pull request characteristics~\cite{li2025riseaiteammatessoftware}, review activity~\cite{bosu2015characteristics}, and reviewer experience~\cite{mcintosh2016empirical}. Following this work, we adopt pull request characteristics, review activity, and participant experience as our explanatory factors, listed in Table~\ref{tab:rq3-metrics} and grouped as follows:

\subsubsection{Pull Request Characteristics}
Pull request characteristics capture the purpose and initial scope of a code change before review begins, including pull request type, initial code churn, initial files changed, and initial commit count~(Table~\ref{tab:rq3-metrics}).

\subsubsection{Review activity} The review activity in the pull request indicates the amount of revision and discussion during the review process. We measure it using commit count, unique reviewer count, and the number of reviewer discussion threads opened on specific code changes.

\subsubsection{Participant experience} Participant experience captures the pull request author and reviewers' familiarity with the project before the current review. As projects now adopt LLM and AI agent reviewers, we also include reviewer type~(e.g., LLM reviewer) as a factor that could explain review quality.

\section{Results}\label{sec:results}


\subsection{RQ1: \rqone}\label{sec:RQ1}

\motivation
With the rapid adoption of AI in software engineering~\cite{hassan2025agenticSE}, generative AI reviewers now participate alongside reviewers across millions of open-source projects~\cite{li2025riseaiteammatessoftware}. Yet, it remains unclear how AI reviewers are introduced and used in software projects over time and how their adoption influences the code review quality. In this research question, we aim to identify common AI reviewer adoption practices and measure their effects on review quality. Our findings provide guidance for projects considering the adoption of AI reviewers.

\approach 
\textbf{Identifying AI Adoption Practices.}\label{subsec:ai_adoption_time_series}
To capture the adoption of LLM and AI agent reviewers in each project, we first construct a time series of the proportion of pull requests reviewed by each reviewer type (e.g., human or AI agent reviewer). Furthermore, similar to prior work~\cite{ReviewBots2020}, to reduce noise caused by AI reviewer participation varying from one pull request to the next, we aggregate the monthly proportion of pull requests reviewed by the era-specific generative AI reviewer type using \emph{era-specific AI reviewer rate} (\(P_{m,e}\)), which is defined as follows:
\begin{equation}
P_{m,e} =
\begin{cases}
0\quad\text{(no GenAI reviewer)},
& e = \text{Pre-LLM}, \\[4pt]
\left(N_m^{\text{LLM}} / N_m\right) \times 100\%,
& e = \text{LLM}, \\[4pt]
\left(N_m^{\text{Agent}} / N_m\right) \times 100\%,
& e = \text{Agent}.
\end{cases}
\label{eq:prior-tech-rate}
\end{equation}

\textit{where \(N_m\) is the total number of pull requests in month \(m\), \(N_m^{\text{LLM}}\) and \(N_m^{\text{Agent}}\) are the numbers of pull requests in month \(m\) involving LLM and AI agent reviewers, respectively.}

To make the AI adoption time series comparable across projects, we use linear interpolation to resample the monthly \(P_{m,e}\) values in each era into 10 evenly spaced points, following prior work on time-series standardization~\cite{friedman1962interpolation,noei2025empirical}. We then concatenate the three eras' normalized segments in chronological order to form the final AI adoption time series. We cluster the resulting time series using soft-DTW clustering~\cite{senin2008dynamic,softdtw2017,noei2025empirical}. Soft-DTW allows time series with similar overall trajectories to be grouped even when corresponding changes occur at slightly different time points. 
To select the number of clusters, we calculate the silhouette score using the same soft-DTW dissimilarity~\cite{rousseeuw1987silhouettes} for different numbers of clusters. The silhouette score ranges from -1 to 1, with a higher value indicating better-separated clusters. We find three as the optimal number of clusters, with a silhouette score of \(0.40\), suggesting fair separation~\cite{silhoutescore}.

\textbf{Quantifying Pull Request Types.} To examine whether LLM and AI agent reviewers are used differently across pull request types, we use the \textit{relative participation rate} metric. The relative participation rate is computed as follows:
\vspace{-2pt}
\begin{equation}
\text{Relative participation rate}_{t} =
\frac{
N_{t}^{\text{AI}} /
\left(N_{t}^{\text{AI}} + N_{t}^{\text{non-AI}}\right)
}{
N_{\text{other}}^{\text{AI}} /
\left(N_{\text{other}}^{\text{AI}} + N_{\text{other}}^{\text{non-AI}}\right)
}
\vspace{-2pt}
\end{equation}

\textit{where \(N_{t}^{\text{AI}}\) and \(N_{t}^{\text{non-AI}}\) denote pull requests of type \(t\) reviewed with and without an LLM or AI agent reviewer, respectively. \(N_{\text{other}}^{\text{AI}}\) and \(N_{\text{other}}^{\text{non-AI}}\) denote pull requests of all other types with and without these reviewers.}

This metric compares the AI reviewer participation rate for type-\(t\) pull requests against that for all other types. We then use a chi-square test~\cite{Hahn01101984}, which measures the associations between categorical variables, to determine whether AI reviewer participation differs significantly across pull request types.

\textbf{Comparing Review Quality.} To examine differences in review quality across eras within each AI adoption practice, we compare the review efficiency and review smell metrics defined in Section~\ref{sec:review-quality-evaluation}. Specifically, the pre-LLM era captures code review practices before generative AI reviewers are adopted, so we use it as the baseline for comparing review efficiency and the prevalence of each review smell in the LLM and agent eras. Since the metric values do not follow a normal distribution, we use the Wilcoxon signed-rank test with Bonferroni correction~\cite{Hahn01101984} for significance testing.


\begin{figure}
    \centering \includegraphics[width=0.95\linewidth]
    {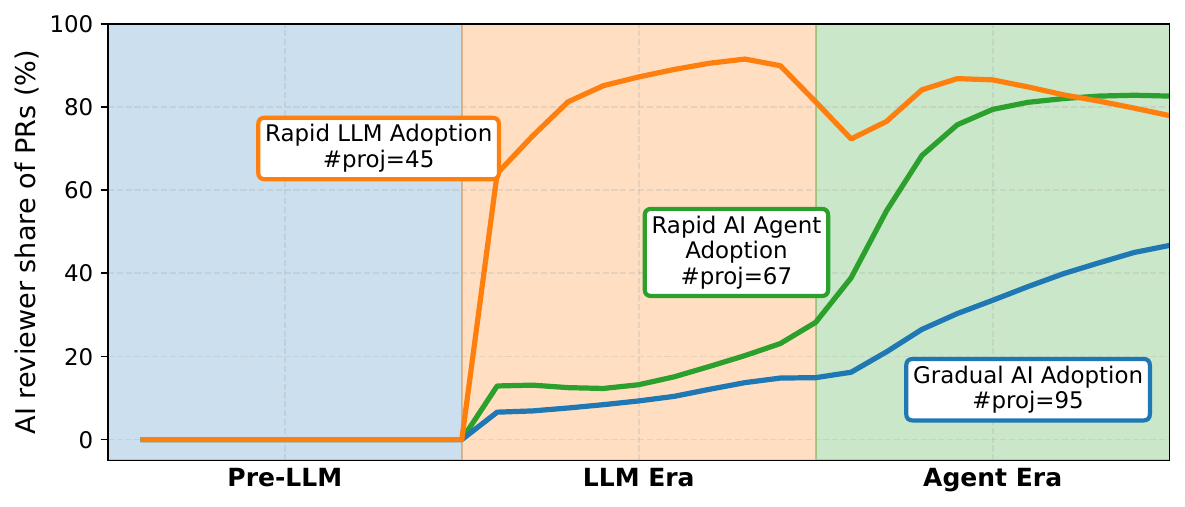}
    \vspace{-8pt}
    \caption{Three identified AI reviewer adoption practices as code review shifts from human-centric to LLM-assisted and agentic review. Higher trend lines indicate greater AI reviewer participation.}
    \vspace{-15pt}
    \label{fig:rq1_clusters}
\end{figure}

\findings \textbf{We identify three AI adoption practices: Gradual AI Adoption~(46\% of 207 studied projects), Rapid LLM Adoption~(22\%), and Rapid AI Agent Adoption~(32\%).} Fig.~\ref{fig:rq1_clusters} illustrates the three identified AI reviewer adoption practices. Below, we describe each practice.
\begin{itemize}
    \item \textit{Gradual AI Adoption} shows a steady transition from human-centric review to generative AI reviewer participation. LLM reviewers participate in only 8\% of pull requests on average in the LLM era, whereas AI agent reviewers participate in 36\% of pull requests in the agent era. The projects span diverse domains, from operating systems~(e.g., RT-Thread~\cite{repo_rtthread}) to educational platforms~(e.g., Archive's OpenLibrary~\cite{repo_openlibrary}), with no single domain dominating.

    \item \textit{Rapid LLM Adoption} captures projects that adopt generative AI reviewers early, with high LLM reviewer participation~(91\% of pull requests on average) in the LLM era followed by high AI agent reviewer participation~(93\% of pull requests on average) in the agent era. In this practice, web applications make up a larger proportion of projects, accounting for 12 of the 45 (27\%).

    \item \textit{Rapid AI Agent Adoption} shows limited LLM reviewer participation in the LLM era, followed by rapid AI agent reviewer uptake in the agent era. In this practice, LLM reviewers participate in 19\% of pull requests on average during the LLM era, followed by AI agent reviewer participation in 76\% of pull requests in the agent era. This practice contains projects from large companies such as Microsoft~\cite{repo_azurecli}~(12 projects) and Google~\cite{repo_flutter}~(3 projects).
\end{itemize}

\begin{table}[!t]
  \centering
  \caption{Review quality and PR types across review eras. We report the smells whose change is statistically significant; smell results are available in the replication package~\cite{github_ai_review_evolution}}
  \vspace{-5pt}
  \label{tab:rq1_combined}
  \setlength{\tabcolsep}{3pt}
  \resizebox{\columnwidth}{!}{%
  \begin{tabular}{c@{\hskip 2pt}lc r r@{}l r@{}l r@{}l @{\hskip 6pt} l}
    \toprule
    \textbf{} & \textbf{} & \textbf{Era} & \textbf{W/AI} & \multicolumn{2}{c}{\textbf{Eff.}} & \multicolumn{2}{c}{\textbf{Smell\%}} & \multicolumn{2}{c}{\textbf{RevBud\%}} & \textbf{PR Types} \\
    \midrule
    \multirow{3}{*}{\rotatebox[origin=c]{90}{\textbf{Gradual}}} & \multirow{3}{*}{\rotatebox[origin=c]{90}{\textbf{AI}}} & PreLLM & --- & 8.5 &  & 83.8 &  & 47.8 &  & --- \\
     &  & LLM & 8\% & -1.6 &  & +3.8 &  & +15.1 &  & chore{\color{green!50!black}\textuparrow{}} fix{\color{green!50!black}\textuparrow{}} \\
     &  & Agent & 36\% & \textcolor{green!60!black}{-2.5} & \textcolor{green!60!black}{*} & +2.0 &  & +15.6 &  & feat{\color{green!50!black}\textuparrow{}} refac{\color{green!50!black}\textuparrow{}} fix{\color{green!50!black}\textuparrow{}} \\
    \midrule
    \multirow{3}{*}{\rotatebox[origin=c]{90}{\textbf{Rapid}}} & \multirow{3}{*}{\rotatebox[origin=c]{90}{\textbf{LLM}}} & PreLLM & --- & 5.4 &  & 83.6 &  & 45.9 &  & --- \\
     &  & LLM & 91\% & +0.6 &  & \textcolor{red!70!black}{+8.0} & \textcolor{red!70!black}{*} & \textcolor{red!70!black}{+26.0} & \textcolor{red!70!black}{*} & docs{\color{green!50!black}\textuparrow{}} build{\color{green!50!black}\textuparrow{}} perf{\color{green!50!black}\textuparrow{}} style{\color{green!50!black}\textuparrow{}} \\
     &  & Agent & 93\% & -0.1 &  & \textcolor{red!70!black}{+4.4} & \textcolor{red!70!black}{*} & \textcolor{red!70!black}{+23.2} & \textcolor{red!70!black}{*} & perf{\color{green!50!black}\textuparrow{}} \\
    \midrule
    \multirow{3}{*}{\rotatebox[origin=c]{90}{\textbf{Rapid}}} & \multirow{3}{*}{\rotatebox[origin=c]{90}{\textbf{Agent}}} & PreLLM & --- & 10.4 &  & 81.0 &  & 36.4 &  & --- \\
     &  & LLM & 19\% & -3.4 &  & +4.4 &  & \textcolor{red!70!black}{+16.6} & \textcolor{red!70!black}{*} & perf{\color{green!50!black}\textuparrow{}} refac{\color{green!50!black}\textuparrow{}} fix{\color{green!50!black}\textuparrow{}} test{\color{green!50!black}\textuparrow{}} feat{\color{green!50!black}\textuparrow{}} \\
     &  & Agent & 76\% & \textcolor{green!60!black}{-4.5} & \textcolor{green!60!black}{*} & +2.5 &  & +12.4 &  & --- \\
    \bottomrule
  \end{tabular}}

  \vspace{-15pt}
\end{table}

\textbf{Gradual AI Adoption and Rapid AI Agent Adoption show significantly more efficient reviews in the agent era.}
Table~\ref{tab:rq1_combined} shows the review efficiency and review smell rate of each practice per era. From the pre-LLM to the agent era, review time drops by $2.5$ days/KLOC for \textit{Gradual AI Adoption}, and by $4.5$ days/KLOC for \textit{Rapid AI Agent Adoption}. However, \textit{Rapid LLM Adoption} shows no significant change across review eras. This suggests that early, heavy reliance on LLM reviewers does not necessarily improve review efficiency, whereas projects that gradually adopt AI reviewers or shift rapidly in the agent era show significant efficiency gains.

\textbf{Rapid LLM Adoption Adoption have a significantly higher review smell rate than their pre-LLM baseline, for both the LLM and agent eras.} As listed in Table~\ref{tab:rq1_combined}, review smells significantly increase in \textit{Rapid LLM Adoption}, by $8.0\%$ points in the LLM era and $4.4\%$ points in the agent era. This increased smell rate is especially reflected in Review Buddies (i.e., repeated assignment of the same reviewer), which rises by 26\% in the LLM era and 23.2\% in the agent era from the pre-LLM baseline. Other review smell types did not show significant increases in our full smell analysis. These results suggest that \textit{Rapid LLM Adoption} may increase reliance on the same LLM reviewer accounts~(e.g., LlamaPReview~\cite{llamapreview}), which narrows review perspectives and reduces feedback diversity.

\textbf{In Gradual AI Adoption, LLM reviewers are utilized significantly more in reviewing maintenance (chore) and bug-fixing pull requests, whereas AI agent reviewers participate significantly more often in feature or refactoring pull requests.} For this adoption practice, LLM reviewers participate more often in chore pull requests~(2.0 times the rate of other types) and fix pull requests~(1.2 times) during the LLM era. In contrast, during the agent era, AI agent reviewers participate more often in feature pull requests~(1.4 times) and refactoring pull requests~(1.3 times). This pattern suggests that, with the gradual shift toward agentic code review, AI-reviewed tasks move from routine maintenance and bug fixing toward more context-aware work such as feature development and refactoring. In Rapid LLM Adoption, generative AI reviewer participation is spread across several pull request types in the LLM era and is only concentrated in performance pull requests in the agent era. In Rapid AI Agent Adoption, generative AI reviewer participation is concentrated in several pull request types during the LLM era but does not concentrate on any single type in the agent era.

\vspace{-5pt}
\smallskip
\begin{Summary}{Summary for RQ1}{}
We identify three AI reviewer adoption practices: \textit{Gradual AI Adoption}, \textit{Rapid LLM Adoption}, and \textit{Rapid AI Agent Adoption}. \textit{Rapid LLM Adoption} significantly increases review smells, whereas \textit{Gradual AI Adoption} and \textit{Rapid AI Agent Adoption} significantly improve review efficiency in the agent review era.
\end{Summary}
\vspace{-5pt} 
\subsection{RQ2: \rqtwo}\label{sec:RQ2}

\motivation
The AI adoption practices capture how projects adopt AI reviewers across review eras. However, they do not show how human, LLM, and AI agent reviewers interact when reviewing individual pull requests. For example, an AI reviewer may initiate the review, or an AI reviewer may join the code review only after human reviewers initiate the discussion. Various interaction sequences can frequently re-occur and therefore emerge as collaboration patterns. In this research question, we aim to identify collaboration patterns and examine their impact on review quality. Understanding these collaboration patterns can guide developers in deciding when and how to involve human, LLM, and AI agent reviewers during code review.

\approach To analyze the collaboration patterns between human and AI reviewers, we represent each pull request as an ordered sequence of review comments labeled by reviewer type~(e.g., AI agent reviewer). We then group these sequences into collaboration patterns and compare their associations with review efficiency and quality across different review eras.

\textbf{Identifying Collaboration Patterns.} To capture reviewer participation within each pull request, we record the temporal sequence of reviewer types from pull request-level review comments and the review decision of the pull request (i.e., accepted or rejected). We order the comments by timestamp and record the reviewer type~(e.g., LLM reviewer) of each comment, preserving repeated participation by the same reviewer type, and end each sequence with the pull request decision. For example, \texttt{Human $\rightarrow$ Agent $\rightarrow$ Human $\rightarrow$ accepted} represents a pull request reviewed first by a human reviewer, then by an AI agent reviewer, then by a human reviewer again before acceptance.

We then abstract reviewer interaction sequences into collaboration patterns using Markov chains with expectation-maximization, following prior software engineering process analyses~\cite{moon1996expectation,song2008trace,bose2009context}. A Markov chain represents one candidate collaboration pattern by estimating the probability of the first reviewer category and the transition probabilities between reviewer categories. Expectation-maximization fits multiple Markov chains to the reviewer interaction sequences and assigns each pull request sequence to the chain that best explains its reviewer transitions. To select the number of collaboration patterns, we fit models with different candidate numbers of Markov chains and compare them using the Bayesian Information Criterion~(BIC)~\cite{watanabe2013widely}. BIC is a model-selection criterion that balances goodness of fit against model complexity, with lower BIC values indicating a better-fitting model. The lowest-BIC model yields 10 collaboration patterns, which we use in the subsequent analysis.

\textbf{Measuring Review Quality.} To measure the quality and efficiency of AI participation in the code review process, we compare the collaboration patterns involving generative AI against human-only code review.

(1) \textit{Review Efficiency:} Within each adoption pattern and era, we use the Scott-Knott Effect Size Difference (ESD) test~\cite{jelihovschi2014scottknott, ouf2026empirical} to rank collaboration patterns by review efficiency. The Scott-Knott ESD test ranks collaboration patterns by significant differences in review efficiency, allowing us to identify the most efficient patterns within the same practice and era.

(2) \textit{Review Smells:} We compare the prevalence of review smells between each AI-involved and human-only collaboration pattern during each combination of adoption practice and era, using chi-square tests with Bonferroni correction~\cite{Hahn01101984}, since smell presence is categorical.

\textbf{Comparing Pull Request Types.} To identify whether collaboration patterns are associated with different pull request types~(e.g., bug fixing), we compare the distribution of the pull request types defined in Table~\ref{tab:pr_types} between each AI-involved collaboration pattern and human-only review. Since collaboration patterns and pull request types are categorical, we use chi-square tests with Bonferroni correction~\cite{Hahn01101984} within each adoption practice and era.

\begin{table}[t]
\centering
\caption{Prevalence of each human-AI review collaboration pattern, its review efficiency rank, review smell prevalence, and the pull request types associated with each pattern. Effi: Scott-Knott ESD rank result for review efficiency; R1 = most efficient. Smell (\%): percentage of pull requests with at least one review smell. PR Type: for Human-Only rows, the top-1 common pull request types with their percentages~(e.g., feat(46) = 46\% are feature pull requests). For other rows, pull request types that are significantly more frequent than in Human-Only reviews within the same era.}
\vspace{-5pt}
\label{tab:rq2_patterns_v2}
\footnotesize
\setlength{\tabcolsep}{0.6pt}
\resizebox{\columnwidth}{!}{
\begin{tabular}{c @{\hskip 1pt} l|c@{\hskip 1pt}c@{\hskip 1pt}l|c@{\hskip 1pt}c@{\hskip 1pt}l|c@{\hskip 1pt}c@{\hskip 1pt}l}
\toprule
\multirow{2}{*}{} & \multirow{2}{*}{\textbf{\shortstack{Collaboration\\Pattern}}}& \multicolumn{3}{c|}{\textbf{Gradual AI Adoption}} & \multicolumn{3}{c|}{\textbf{Rapid LLM Adoption}} & \multicolumn{3}{c}{\textbf{Rapid Agent Adoption}}  \\
\cline{3-5}\cline{6-8}\cline{9-11}
 &  & \shortstack{Effi.} & \shortstack{Smell} & \multicolumn{1}{l|}{ PR Type} & \shortstack{Effi.} & \shortstack{Smell} & \multicolumn{1}{l|}{ PR Type} & \shortstack{Effi.} & \shortstack{Smell} & \multicolumn{1}{l}{ PR Type} \\
\midrule
\multirow{3}{*}{\rotatebox[origin=c]{90}{\textbf{Pre-LLM}}} & Human-Only & {\normalsize\colorbox{rank2}{R2}} & 75 & feat(46) & {\normalsize\colorbox{rank2}{R2}} & 72 & feat(46) & {\normalsize\colorbox{rank2}{R2}} & 75 & feat(43) \\
 & Human-Bot & {\normalsize\colorbox{rank2}{R2}} & 84{\color{red!70!black}\textuparrow{}} & chore{\color{green!50!black}\textuparrow{}} & {\normalsize\colorbox{rank1}{R1}} & 86{\color{red!70!black}\textuparrow{}} & & {\normalsize\colorbox{rank2}{R2}} & 84{\color{red!70!black}\textuparrow{}} & \\
 & Human-Bot-ML & {\normalsize\colorbox{rank1}{R1}} & 88{\color{red!70!black}\textuparrow{}} & chore{\color{green!50!black}\textuparrow{}} fix{\color{green!50!black}\textuparrow{}} & {\normalsize\colorbox{rank1}{R1}} & 82{\color{red!70!black}\textuparrow{}} & chore{\color{green!50!black}\textuparrow{}} fix{\color{green!50!black}\textuparrow{}} & {\normalsize\colorbox{rank1}{R1}} & 87{\color{red!70!black}\textuparrow{}} & feat{\color{green!50!black}\textuparrow{}} \\
\hline
\multirow{5}{*}{\rotatebox[origin=c]{90}{\textbf{LLM}}} & Human-Only & {\normalsize\colorbox{rank1}{R1}} & 76 & feat(41) & {\normalsize\colorbox{rank1}{R1}} & 69 & chore(32) & {\normalsize\colorbox{rank1}{R1}} & 72 & feat(33)\\
 & Human-Bot & {\normalsize\colorbox{rank1}{R1}} & 82{\color{red!70!black}\textuparrow{}} & chore{\color{green!50!black}\textuparrow{}} & {\normalsize\colorbox{rank2}{R2}} & 78{\color{red!70!black}\textuparrow{}} & fix{\color{green!50!black}\textuparrow{}} & {\normalsize\colorbox{rank1}{R1}} & 88{\color{red!70!black}\textuparrow{}} & \\
 & LLM-Assist & {\normalsize\colorbox{rank2}{R2}} & 81{\color{red!70!black}\textuparrow{}} & & {\normalsize\colorbox{rank3}{R3}} & 87{\color{red!70!black}\textuparrow{}} & fix{\color{green!50!black}\textuparrow{}} & {\normalsize\colorbox{rank2}{R2}} & 83{\color{red!70!black}\textuparrow{}} & feat{\color{green!50!black}\textuparrow{}} \\
 & Multi-LLM & {\normalsize\colorbox{rank1}{R1}} & 81 & chore{\color{green!50!black}\textuparrow{}} feat{\color{green!50!black}\textuparrow{}} & {\normalsize\colorbox{rank2}{R2}} & 86{\color{red!70!black}\textuparrow{}} & feat{\color{green!50!black}\textuparrow{}} fix{\color{green!50!black}\textuparrow{}} & {\normalsize\colorbox{rank1}{R1}} & 81{\color{red!70!black}\textuparrow{}} & fix{\color{green!50!black}\textuparrow{}} \\
 & LLM-Bot-Assist & {\normalsize\colorbox{rank1}{R1}} & 94{\color{red!70!black}\textuparrow{}} & chore{\color{green!50!black}\textuparrow{}} refac{\color{green!50!black}\textuparrow{}} & {\normalsize\colorbox{rank2}{R2}} & 80{\color{red!70!black}\textuparrow{}} & docs{\color{green!50!black}\textuparrow{}} fix{\color{green!50!black}\textuparrow{}} & {\normalsize\colorbox{rank2}{R2}} & 87{\color{red!70!black}\textuparrow{}} & fix{\color{green!50!black}\textuparrow{}} refac{\color{green!50!black}\textuparrow{}} \\
\hline
\multirow{8}{*}{\rotatebox[origin=c]{90}{\textbf{Agent}}} & Human-Only & {\normalsize\colorbox{rank3}{R3}} & 75 & feat(42) & {\normalsize\colorbox{rank1}{R1}} & 74 & feat(34) & {\normalsize\colorbox{rank2}{R2}} & 71 & feat(39) \\
 & Human-Bot & {\normalsize\colorbox{rank2}{R2}} & 83{\color{red!70!black}\textuparrow{}} & chore{\color{green!50!black}\textuparrow{}} & {\normalsize\colorbox{rank2}{R2}} & 73 & fix{\color{red!70!black}\textdownarrow{}} docs{\color{green!50!black}\textuparrow{}}& {\normalsize\colorbox{rank2}{R2}} & 84{\color{red!70!black}\textuparrow{}} & \\
 & Agent-Init & {\normalsize\colorbox{rank2}{R2}} & 78{\color{red!70!black}\textuparrow{}} & chore{\color{green!50!black}\textuparrow{}} & {\normalsize\colorbox{rank1}{R1}} & 84{\color{red!70!black}\textuparrow{}} & fix{\color{red!70!black}\textdownarrow{}} chore{\color{green!50!black}\textuparrow{}} & {\normalsize\colorbox{rank1}{R1}} & 80{\color{red!70!black}\textuparrow{}} & chore{\color{green!50!black}\textuparrow{}} \\
 & Agent-Assist & {\normalsize\colorbox{rank3}{R3}} & 84{\color{red!70!black}\textuparrow{}} & & {\normalsize\colorbox{rank2}{R2}} & 84{\color{red!70!black}\textuparrow{}} & fix{\color{red!70!black}\textdownarrow{}} feat{\color{green!50!black}\textuparrow{}} & {\normalsize\colorbox{rank2}{R2}} & 83{\color{red!70!black}\textuparrow{}} & \\
 & Multi-Agent & {\normalsize\colorbox{rank1}{R1}} & 83{\color{red!70!black}\textuparrow{}} & chore{\color{green!50!black}\textuparrow{}} & {\normalsize\colorbox{rank1}{R1}} & 83{\color{red!70!black}\textuparrow{}} & fix{\color{red!70!black}\textdownarrow{}} docs{\color{green!50!black}\textuparrow{}} & {\normalsize\colorbox{rank1}{R1}} & 82{\color{red!70!black}\textuparrow{}} & fix{\color{green!50!black}\textuparrow{}} \\
 & Agent-ML-Assist & {\normalsize\colorbox{rank3}{R3}} & 86{\color{red!70!black}\textuparrow{}} & & {\normalsize\colorbox{rank2}{R2}} & 84{\color{red!70!black}\textuparrow{}} & feat{\color{green!50!black}\textuparrow{}} & {\normalsize\colorbox{rank2}{R2}} & 84{\color{red!70!black}\textuparrow{}} & \\

\bottomrule
\end{tabular}
}
\vspace{-15pt}
\end{table}

\findings \textbf{In the LLM era, human-bot and multi-LLM collaboration patterns are as efficient as human-only review in Gradual AI Adoption and Rapid AI Agent Adoption. However, for Rapid LLM Adoption, all LLM-involved review patterns are significantly slower than human-only review.} Table~\ref{tab:rq2_patterns_v2} shows the Scott-Knott ESD efficiency rank for collaboration patterns within each adoption practice and review era. As listed in Table~\ref{tab:rq2_patterns_v2}, LLM-Assist reviews are less efficient than human-only reviews across all three adoption practices~(R2--R3 vs. R1). Multi-LLM reviews are as efficient as human-only reviews under Gradual AI Adoption and Rapid AI Agent Adoption~(both R1), but are less efficient under Rapid LLM Adoption~(R2 vs. R1). In Multi-LLM reviews, the LLM reviewer contributes a median of two comments, and in 75\% of these pull requests, human reviewers respond with only a single comment before the review ends. By contrast, in LLM-Assist reviews, humans comment first and often continue the discussion after the LLM comment, resulting in more human back-and-forth than in human-only reviews~(with medians of four and two human comments, respectively; see our replication package~\cite{github_ai_review_evolution}). This suggests that a single LLM comment is often insufficient to resolve a pull request by itself. Instead, reviewers still need continued human-AI back-and-forth before reaching a decision.

\textbf{In the agent era, agent-init and multi-agent reviews are significantly faster than human-only review under Gradual AI and Rapid AI Agent Adoption.} As listed in Table~\ref{tab:rq2_patterns_v2}, agent-initiated and multi-agent reviews rank ahead of human-only review (R1–R2 vs. R2–R3) for both Gradual AI and Rapid AI Agent Adoption. However, we do not observe a significant efficiency difference from human-only review under Rapid LLM Adoption. In the agent era of Rapid LLM Adoption, human-only review ranks first in review efficiency, with 61\% of its pull requests resolved in a two-turn exchange~\cite{github_ai_review_evolution}. This indicates that human review is already fast in this practice. Furthermore, we observe that in agent-init and multi-agent reviews, the agents take over the inspection step in 75\% to 95\% of these pull requests, posting a summary of the code change, after which the human responds with a single short comment. For example, after the agent summarizes a bug fix, the human replies \textit{``I cannot reproduce the bug, but the change looks very reasonable to me.''}, after which the pull request is merged. This suggests that agent reviewers can support the inspection step and provide a useful starting point for the review, potentially reducing the exchanges needed from human reviewers.

\textbf{Pull requests reviewed only by human reviewers exhibit significantly lower review smell prevalence than most AI-involved collaboration patterns.} Table~\ref{tab:rq2_patterns_v2} shows that smell prevalence ranges from 69\% to 76\% for human-only reviews and from 78\% to 94\% for patterns involving AI across the LLM and agent eras. This higher smell prevalence is mainly driven by Review Buddies, which rises from 16\% for human-only reviews to 60\% for LLM-involved patterns and 53\% for agent-involved patterns on average~\cite{github_ai_review_evolution}. The other smells do not follow this trend, with Sleeping Review declining and Large Changeset staying flat across eras~\cite{github_ai_review_evolution}. These results suggest that repeated use of the same AI reviewer or model identity may reduce reviewer diversity and narrow review perspectives. Practitioners could adopt more collaborative multi-agent setups that distribute review across several agents and encourage knowledge transfer, so that the review process does not depend on a single default AI reviewer.

\textbf{AI-involved collaboration patterns are associated with different pull request types across adoption practices and eras.} As shown in Table~\ref{tab:rq2_patterns_v2}, under Gradual AI Adoption, several AI-involved patterns, such as Multi-LLM, LLM-Bot-Assist, and Multi-Agent, are associated with chore pull requests more often than human-only review. Under Rapid LLM Adoption, most AI-involved patterns in the LLM era are associated with bug-fixing pull requests and are less efficient than human-only review~(R2--R3 vs. R1). In the agent era of Rapid LLM Adoption, fix pull requests become significantly less frequent for several AI-involved patterns, such as Agent-Init and Multi-Agent, while other patterns are associated with documentation, chore, or feature pull requests. These results suggest that, under Rapid LLM Adoption, agent-era collaboration patterns change the types of pull requests involving AI more than they change review efficiency.

\smallskip
\vspace{-5pt}
\begin{Summary}{Summary for RQ2}{}
Collaboration patterns with LLM or AI agent reviewers show significantly higher review buddies than human-only review. In the agent era, agent-init and multi-agent reviews are significantly faster than human-only review in Gradual AI and Rapid AI Agent Adoption, but show no significant difference in Rapid LLM Adoption.
\end{Summary}
\vspace{-5pt}
\subsection{RQ3: \rqthree}\label{sec:RQ3}

\motivation In RQ1 and RQ2, we find that review quality differs significantly across AI adoption practices and human-AI collaboration patterns. However, this quality difference may not necessarily result from AI collaboration patterns or AI adoption practices, and may be explained by other factors of the review process, such as reviewer experience and review activity~(e.g., commit count, inline discussion threads) within a pull request~\cite{thongtanunam2017review,baysal2016investigating}. We therefore examine whether review quality is more strongly associated with collaboration patterns or with alternative factors in the review process~(e.g., pull request characteristics)~. This analysis helps identify the factors associated with better review quality, guiding what developers could consider when designing agentic review pipelines.

\approach To examine how traditional review factors and human-AI collaboration factors are associated with review quality, we build explanatory models for each AI adoption practice and review era, using the following process.

\textbf{Training Explanatory Model.} To examine how our explanatory factors are associated with review quality, we use logistic regression~\cite{tourani2016impact, ghaleb2019studying,justintime2013TSE}. Logistic regression estimates how each factor is associated with the probability of an efficient review or the presence of a review smell while accounting for the other factors in the model. In each model, the independent variables include the collaboration patterns~(identified in Sec.~\ref{sec:RQ2}), pull request characteristics~(e.g., initial code churn), review activity~(e.g., commit counts during the review), and participant experience(e.g., human reviewer experience in the same project), which are listed in Table~\ref{tab:rq3-metrics}. The dependent variable is a binary review outcome that captures whether a review is efficient or whether it contains a given review smell, which is labeled as follows:

\textit{(1) Review efficiency analysis:} The review efficiency metric is continuous~(\#days/KLOC), so we discretize it into ``efficient'' and ``inefficient'' classes. Within each adoption practice and review era, pull requests below the median \#days/KLOC are labeled ``efficient'', and the rest are labeled ``inefficient''. This yields nine efficiency models (i.e., 3 practices \(\times\) 3 eras) in total, which estimate whether the explanatory variables are associated with efficient review.

\textit{(2) Review smells analysis:} For each selected review smell, the dependent variable is binary, indicating whether a pull request contains that smell. Because the review smell models are trained separately for each adoption practice and review era, each modeled smell must include enough pull requests with and without that smell in every model. Therefore, we focus on the three most prevalent smells among modeled pull requests: Review Buddies~(46.9\%), Sleeping Review~(38.6\%), and Large Changeset~(20.6\%). This yields 27 smell models in total~(3 smells \(\times\) 3 practices \(\times\) 3 eras), which estimate whether the explanatory variables are associated with each smell.

\textbf{Processing explanatory variables.} Since highly correlated variables can distort model estimates~\cite{ghaleb2019studying,noei2025detecting}, we compute the Variance Inflation Factor~(VIF) for each variable in Table~\ref{tab:rq3-metrics}. VIF measures how much a variable's variance is inflated by its correlation with the other variables~\cite{o2007caution}. All variables yielded a VIF below 5~\cite{tourani2016impact}, so none are removed. To reduce the influence of extreme values, we then apply a log transformation to each numeric variable with positive skew~\cite{justintime2013TSE}. For categorical variables~(e.g., collaboration pattern and pull request types), we use dummy encoding. For collaboration patterns, human-only review serves as the reference category. For pull request type, the ``feature'' type serves as the reference category. Thus, coefficients for dummy-encoded collaboration patterns are interpreted relative to human-only review, and those for pull request types are interpreted relative to feature pull requests~\cite{tourani2016impact}.

\textbf{Measuring the goodness of fit.}
To analyze how well the explanatory models distinguish between the binary review quality classifications, we use the Area Under the ROC Curve (AUC)~\cite{bradley1997use, yu2026empirical} to assess whether the explanatory models provide sufficient separation between the binary review outcomes used in our analysis. An AUC of 0.5 indicates random separation, while substantially larger (or smaller) values indicate that the explanatory variables better distinguish efficient reviews from inefficient ones or smelly from non-smelly reviews. To avoid evaluation on the same data used to train the model, we train each model (9+27=36 models in total) on 90\% of the data within each combination of AI adoption practice and review era, and use the remaining 10\% held-out test set to measure AUC~\cite{tourani2016impact}.

\begin{table*}[!t]
\centering
\caption{Impact of collaboration patterns and traditional factors on review delay (lack of efficiency) and review smells across review eras and AI-adoption practices. Green text indicates a significantly better outcome (e.g., lower review delay), whereas red text indicates a significantly worse outcome. Blank cells indicate non-significant associations~($p>=0.05$ and $|\mathrm{impact}|< 10\%$.).}
\label{tab:rq3_effi_smell_byera}

\vspace{-5pt}
\footnotesize
\setlength{\tabcolsep}{2pt}
\renewcommand{\arraystretch}{1.1}
\definecolor{goodgreen}{HTML}{1A9850}
\definecolor{badred}{HTML}{D73027}
\resizebox{\textwidth}{!}{%
\begin{tabular}{!{\vrule width 1.28pt}c|l!{\vrule width 1.28pt}c!{\vrule width 0.1pt}c!{\vrule width 0.1pt}c!{\vrule width 1.12pt}c!{\vrule width 0.1pt}c!{\vrule width 0.1pt}c!{\vrule width 1.12pt}c!{\vrule width 0.1pt}c!{\vrule width 0.1pt}c!{\vrule width 1.28pt}c!{\vrule width 0.1pt}c!{\vrule width 0.1pt}c!{\vrule width 1.12pt}c!{\vrule width 0.1pt}c!{\vrule width 0.1pt}c!{\vrule width 1.12pt}c!{\vrule width 0.1pt}c!{\vrule width 0.1pt}c!{\vrule width 1.28pt}c!{\vrule width 0.1pt}c!{\vrule width 0.1pt}c!{\vrule width 1.12pt}c!{\vrule width 0.1pt}c!{\vrule width 0.1pt}c!{\vrule width 1.12pt}c!{\vrule width 0.1pt}c!{\vrule width 0.1pt}c!{\vrule width 1.28pt}c!{\vrule width 0.1pt}c!{\vrule width 0.1pt}c!{\vrule width 1.12pt}c!{\vrule width 0.1pt}c!{\vrule width 0.1pt}c!{\vrule width 1.12pt}c!{\vrule width 0.1pt}c!{\vrule width 0.1pt}c!{\vrule width 1.28pt}}
\specialrule{1.28pt}{0pt}{0pt}
\multicolumn{2}{!{\vrule width 1.28pt}c!{\vrule width 1.28pt}}{} & \multicolumn{9}{c!{\vrule width 1.28pt}}{\multirow{2}{*}{\textbf{(1) Lack of Efficiency~(Review Delay)}}} & \multicolumn{27}{c!{\vrule width 1.28pt}}{\textbf{Review Smells}} \\
\cline{12-38}
\multicolumn{2}{!{\vrule width 1.28pt}c!{\vrule width 1.28pt}}{} & \multicolumn{9}{c!{\vrule width 1.28pt}}{} & \multicolumn{9}{c!{\vrule width 1.28pt}}{\textbf{(2) Review Buddies}} & \multicolumn{9}{c!{\vrule width 1.28pt}}{\textbf{(3) Sleeping Review}} & \multicolumn{9}{c!{\vrule width 1.28pt}}{\textbf{(4) Large Changeset}} \\
\cline{3-38}
\multicolumn{2}{!{\vrule width 1.28pt}c!{\vrule width 1.28pt}}{\textbf{Factor}} & \multicolumn{3}{c!{\vrule width 1.12pt}}{\textbf{Pre-LLM}} & \multicolumn{3}{c!{\vrule width 1.12pt}}{\textbf{LLM}} & \multicolumn{3}{c!{\vrule width 1.28pt}}{\textbf{Agent}} & \multicolumn{3}{c!{\vrule width 1.12pt}}{\textbf{Pre-LLM}} & \multicolumn{3}{c!{\vrule width 1.12pt}}{\textbf{LLM}} & \multicolumn{3}{c!{\vrule width 1.28pt}}{\textbf{Agent}} & \multicolumn{3}{c!{\vrule width 1.12pt}}{\textbf{Pre-LLM}} & \multicolumn{3}{c!{\vrule width 1.12pt}}{\textbf{LLM}} & \multicolumn{3}{c!{\vrule width 1.28pt}}{\textbf{Agent}} & \multicolumn{3}{c!{\vrule width 1.12pt}}{\textbf{Pre-LLM}} & \multicolumn{3}{c!{\vrule width 1.12pt}}{\textbf{LLM}} & \multicolumn{3}{c!{\vrule width 1.28pt}}{\textbf{Agent}} \\
\hline
\multirow{7}{*}{\rotatebox[origin=c]{90}{\textbf{Collaboration}}} & \textbf{Multi-Agent} &  &  &  &  &  &  & \textcolor{goodgreen}{\textbf{$-$50}} & \textcolor{badred}{\textbf{$+$37}} & \textcolor{goodgreen}{\textbf{$-$13}} &  &  &  &  &  &  & \textcolor{badred}{\textbf{$+$220}} & \textcolor{badred}{\textbf{$+$250}} & \textcolor{badred}{\textbf{$+$513}} &  &  &  &  &  &  & \textcolor{goodgreen}{\textbf{$-$48}} &  & \textcolor{goodgreen}{\textbf{$-$16}} &  &  &  &  &  &  & \textcolor{badred}{\textbf{$+$29}} & \textcolor{goodgreen}{\textbf{$-$60}} & \textcolor{badred}{\textbf{$+$54}} \\
 & \textbf{Agent-Init} &  &  &  &  &  &  & \textcolor{goodgreen}{\textbf{$-$37}} & \textcolor{badred}{\textbf{$+$39}} & \textcolor{goodgreen}{\textbf{$-$16}} &  &  &  &  &  &  & \textcolor{badred}{\textbf{$+$200}} & \textcolor{badred}{\textbf{$+$249}} & \textcolor{badred}{\textbf{$+$605}} &  &  &  &  &  &  & \textcolor{goodgreen}{\textbf{$-$36}} & \textcolor{badred}{\textbf{$+$42}} & \textcolor{goodgreen}{\textbf{$-$26}} &  &  &  &  &  &  & \textcolor{badred}{\textbf{$+$27}} & \textcolor{goodgreen}{\textbf{$-$55}} & \textcolor{badred}{\textbf{$+$20}} \\
 & \textbf{Agent-ML} &  &  &  &  &  &  & \textcolor{goodgreen}{\textbf{$-$11}} & \textcolor{badred}{\textbf{$+$50}} &  &  &  &  &  &  &  & \textcolor{badred}{\textbf{$+$246}} & \textcolor{badred}{\textbf{$+$269}} & \textcolor{badred}{\textbf{$+$678}} &  &  &  &  &  &  &  & \textcolor{badred}{\textbf{$+$94}} &  &  &  &  &  &  &  & \textcolor{badred}{\textbf{$+$48}} & \textcolor{goodgreen}{\textbf{$-$44}} &  \\
 & \textbf{Agent-Assist} &  &  &  &  &  &  &  & \textcolor{badred}{\textbf{$+$53}} & \textcolor{badred}{\textbf{$+$13}} &  &  &  &  &  &  & \textcolor{badred}{\textbf{$+$174}} & \textcolor{badred}{\textbf{$+$227}} & \textcolor{badred}{\textbf{$+$418}} &  &  &  &  &  &  &  & \textcolor{badred}{\textbf{$+$88}} & \textcolor{badred}{\textbf{$+$26}} &  &  &  &  &  &  & \textcolor{badred}{\textbf{$+$25}} & \textcolor{goodgreen}{\textbf{$-$51}} & \textcolor{badred}{\textbf{$+$29}} \\
 & \textbf{LLM-Bot} &  &  &  &  & \textcolor{badred}{\textbf{$+$33}} & \textcolor{badred}{\textbf{$+$28}} &  &  &  &  &  &  & \textcolor{badred}{\textbf{$+$220}} & \textcolor{badred}{\textbf{$+$2K}} & \textcolor{badred}{\textbf{$+$130}} &  &  &  &  &  &  & \textcolor{badred}{\textbf{$+$26}} & \textcolor{badred}{\textbf{$+$124}} & \textcolor{badred}{\textbf{$+$56}} &  &  &  &  &  &  &  & \textcolor{goodgreen}{\textbf{$-$44}} & \textcolor{goodgreen}{\textbf{$-$41}} &  &  &  \\
 & \textbf{LLM-Assist} &  &  &  & \textcolor{badred}{\textbf{$+$28}} & \textcolor{badred}{\textbf{$+$45}} & \textcolor{badred}{\textbf{$+$30}} &  &  &  &  &  &  & \textcolor{badred}{\textbf{$+$65}} & \textcolor{badred}{\textbf{$+$2K}} & \textcolor{badred}{\textbf{$+$62}} &  &  &  &  &  &  & \textcolor{badred}{\textbf{$+$45}} & \textcolor{badred}{\textbf{$+$214}} & \textcolor{badred}{\textbf{$+$79}} &  &  &  &  &  &  &  & \textcolor{goodgreen}{\textbf{$-$43}} &  &  &  &  \\
 & \textbf{Human-Bot} &  &  &  &  & \textcolor{badred}{\textbf{$+$17}} & \textcolor{badred}{\textbf{$+$12}} & \textcolor{goodgreen}{\textbf{$-$21}} &  &  & \textcolor{badred}{\textbf{$+$287}} & \textcolor{badred}{\textbf{$+$1K}} & \textcolor{badred}{\textbf{$+$283}} & \textcolor{badred}{\textbf{$+$170}} & \textcolor{badred}{\textbf{$+$2K}} & \textcolor{badred}{\textbf{$+$108}} & \textcolor{badred}{\textbf{$+$197}} & \textcolor{badred}{\textbf{$+$96}} & \textcolor{badred}{\textbf{$+$437}} &  & \textcolor{goodgreen}{\textbf{$-$14}} &  &  & \textcolor{badred}{\textbf{$+$46}} & \textcolor{badred}{\textbf{$+$32}} & \textcolor{goodgreen}{\textbf{$-$24}} &  &  & \textcolor{goodgreen}{\textbf{$-$21}} &  &  & \textcolor{goodgreen}{\textbf{$-$18}} & \textcolor{goodgreen}{\textbf{$-$35}} & \textcolor{badred}{\textbf{$+$38}} & \textcolor{badred}{\textbf{$+$14}} & \textcolor{badred}{\textbf{$+$26}} & \textcolor{badred}{\textbf{$+$20}} \\
\hline
\multirow{6}{*}{\rotatebox[origin=c]{90}{\textbf{Traditional}}} & \textbf{build} &  &  & \textcolor{goodgreen}{\textbf{$-$18}} &  &  & \textcolor{goodgreen}{\textbf{$-$20}} & \textcolor{badred}{\textbf{$+$16}} &  & \textcolor{goodgreen}{\textbf{$-$40}} & \textcolor{goodgreen}{\textbf{$-$61}} &  & \textcolor{goodgreen}{\textbf{$-$15}} & \textcolor{goodgreen}{\textbf{$-$86}} & \textcolor{goodgreen}{\textbf{$-$43}} & \textcolor{goodgreen}{\textbf{$-$20}} & \textcolor{goodgreen}{\textbf{$-$41}} & \textcolor{goodgreen}{\textbf{$-$55}} & \textcolor{goodgreen}{\textbf{$-$19}} & \textcolor{goodgreen}{\textbf{$-$10}} & \textcolor{goodgreen}{\textbf{$-$38}} & \textcolor{goodgreen}{\textbf{$-$26}} & \textcolor{goodgreen}{\textbf{$-$26}} & \textcolor{goodgreen}{\textbf{$-$22}} & \textcolor{goodgreen}{\textbf{$-$14}} &  & \textcolor{goodgreen}{\textbf{$-$24}} & \textcolor{goodgreen}{\textbf{$-$25}} &  & \textcolor{goodgreen}{\textbf{$-$31}} &  & \textcolor{goodgreen}{\textbf{$-$47}} & \textcolor{badred}{\textbf{$+$52}} & \textcolor{badred}{\textbf{$+$29}} &  & \textcolor{badred}{\textbf{$+$20}} & \textcolor{badred}{\textbf{$+$44}} \\
 & \textbf{chore} & \textcolor{goodgreen}{\textbf{$-$32}} & \textcolor{goodgreen}{\textbf{$-$41}} & \textcolor{goodgreen}{\textbf{$-$24}} & \textcolor{goodgreen}{\textbf{$-$30}} & \textcolor{goodgreen}{\textbf{$-$11}} & \textcolor{goodgreen}{\textbf{$-$12}} &  &  & \textcolor{goodgreen}{\textbf{$-$25}} & \textcolor{badred}{\textbf{$+$19}} & \textcolor{badred}{\textbf{$+$52}} & \textcolor{badred}{\textbf{$+$27}} &  & \textcolor{goodgreen}{\textbf{$-$39}} & \textcolor{goodgreen}{\textbf{$-$30}} & \textcolor{badred}{\textbf{$+$14}} & \textcolor{goodgreen}{\textbf{$-$25}} &  & \textcolor{goodgreen}{\textbf{$-$40}} & \textcolor{goodgreen}{\textbf{$-$50}} & \textcolor{goodgreen}{\textbf{$-$42}} & \textcolor{goodgreen}{\textbf{$-$48}} & \textcolor{goodgreen}{\textbf{$-$33}} & \textcolor{goodgreen}{\textbf{$-$40}} & \textcolor{goodgreen}{\textbf{$-$14}} & \textcolor{goodgreen}{\textbf{$-$28}} & \textcolor{goodgreen}{\textbf{$-$32}} & \textcolor{badred}{\textbf{$+$31}} &  &  & \textcolor{badred}{\textbf{$+$29}} &  &  &  &  & \textcolor{badred}{\textbf{$+$11}} \\
 & \textbf{refactor} & \textcolor{goodgreen}{\textbf{$-$11}} & \textcolor{goodgreen}{\textbf{$-$16}} & \textcolor{goodgreen}{\textbf{$-$12}} &  & \textcolor{goodgreen}{\textbf{$-$10}} &  & \textcolor{goodgreen}{\textbf{$-$11}} &  &  &  & \textcolor{badred}{\textbf{$+$34}} & \textcolor{badred}{\textbf{$+$29}} & \textcolor{badred}{\textbf{$+$25}} & \textcolor{goodgreen}{\textbf{$-$19}} & \textcolor{badred}{\textbf{$+$20}} & \textcolor{badred}{\textbf{$+$17}} & \textcolor{goodgreen}{\textbf{$-$12}} & \textcolor{badred}{\textbf{$+$33}} &  & \textcolor{goodgreen}{\textbf{$-$22}} & \textcolor{goodgreen}{\textbf{$-$14}} &  & \textcolor{goodgreen}{\textbf{$-$24}} &  &  & \textcolor{goodgreen}{\textbf{$-$19}} &  & \textcolor{badred}{\textbf{$+$32}} & \textcolor{badred}{\textbf{$+$12}} &  & \textcolor{badred}{\textbf{$+$14}} &  & \textcolor{badred}{\textbf{$+$15}} & \textcolor{badred}{\textbf{$+$26}} & \textcolor{badred}{\textbf{$+$15}} &  \\
 & \textbf{Author Exp} & \textcolor{goodgreen}{\textbf{$-$17}} & \textcolor{goodgreen}{\textbf{$-$22}} & \textcolor{goodgreen}{\textbf{$-$18}} & \textcolor{goodgreen}{\textbf{$-$12}} &  &  & \textcolor{goodgreen}{\textbf{$-$15}} &  & \textcolor{goodgreen}{\textbf{$-$14}} & \textcolor{badred}{\textbf{$+$115}} & \textcolor{badred}{\textbf{$+$238}} & \textcolor{badred}{\textbf{$+$151}} & \textcolor{badred}{\textbf{$+$72}} & \textcolor{badred}{\textbf{$+$516}} & \textcolor{badred}{\textbf{$+$67}} & \textcolor{badred}{\textbf{$+$93}} & \textcolor{badred}{\textbf{$+$228}} & \textcolor{badred}{\textbf{$+$116}} & \textcolor{goodgreen}{\textbf{$-$19}} & \textcolor{goodgreen}{\textbf{$-$24}} & \textcolor{goodgreen}{\textbf{$-$22}} & \textcolor{goodgreen}{\textbf{$-$19}} & \textcolor{goodgreen}{\textbf{$-$23}} & \textcolor{goodgreen}{\textbf{$-$24}} & \textcolor{goodgreen}{\textbf{$-$17}} & \textcolor{goodgreen}{\textbf{$-$24}} & \textcolor{goodgreen}{\textbf{$-$18}} &  & \textcolor{badred}{\textbf{$+$12}} & \textcolor{badred}{\textbf{$+$10}} &  & \textcolor{badred}{\textbf{$+$10}} &  & \textcolor{badred}{\textbf{$+$12}} &  & \textcolor{badred}{\textbf{$+$20}} \\
 & \textbf{NumCommit} &  &  &  &  &  &  &  &  &  &  & \textcolor{goodgreen}{\textbf{$-$14}} &  &  &  & \textcolor{badred}{\textbf{$+$10}} &  &  &  & \textcolor{badred}{\textbf{$+$36}} & \textcolor{badred}{\textbf{$+$31}} & \textcolor{badred}{\textbf{$+$28}} & \textcolor{badred}{\textbf{$+$48}} & \textcolor{badred}{\textbf{$+$70}} & \textcolor{badred}{\textbf{$+$41}} & \textcolor{badred}{\textbf{$+$29}} & \textcolor{badred}{\textbf{$+$27}} & \textcolor{badred}{\textbf{$+$37}} & \textcolor{badred}{\textbf{$+$95}} & \textcolor{badred}{\textbf{$+$100}} & \textcolor{badred}{\textbf{$+$73}} & \textcolor{badred}{\textbf{$+$96}} & \textcolor{badred}{\textbf{$+$73}} & \textcolor{badred}{\textbf{$+$77}} & \textcolor{badred}{\textbf{$+$74}} & \textcolor{badred}{\textbf{$+$49}} & \textcolor{badred}{\textbf{$+$76}} \\
 & \textbf{InlineThread} & \textcolor{badred}{\textbf{$+$11}} & \textcolor{badred}{\textbf{$+$20}} & \textcolor{badred}{\textbf{$+$15}} & \textcolor{badred}{\textbf{$+$12}} &  & \textcolor{badred}{\textbf{$+$15}} & \textcolor{badred}{\textbf{$+$12}} &  & \textcolor{badred}{\textbf{$+$10}} &  &  &  &  & \textcolor{badred}{\textbf{$+$16}} & \textcolor{goodgreen}{\textbf{$-$19}} &  &  & \textcolor{goodgreen}{\textbf{$-$15}} & \textcolor{badred}{\textbf{$+$30}} & \textcolor{badred}{\textbf{$+$36}} & \textcolor{badred}{\textbf{$+$30}} & \textcolor{badred}{\textbf{$+$45}} & \textcolor{badred}{\textbf{$+$43}} & \textcolor{badred}{\textbf{$+$51}} & \textcolor{badred}{\textbf{$+$30}} & \textcolor{badred}{\textbf{$+$26}} & \textcolor{badred}{\textbf{$+$27}} & \textcolor{badred}{\textbf{$+$16}} & \textcolor{badred}{\textbf{$+$12}} &  & \textcolor{badred}{\textbf{$+$33}} & \textcolor{badred}{\textbf{$+$44}} &  & \textcolor{badred}{\textbf{$+$32}} & \textcolor{badred}{\textbf{$+$30}} & \textcolor{badred}{\textbf{$+$24}} \\
\specialrule{1.28pt}{0pt}{0pt}
\multicolumn{2}{c}{} & \multicolumn{1}{c}{\makebox[0pt][r]{\rotatebox[origin=rB]{45}{\textbf{Gradual AI}}}} & \multicolumn{1}{c}{\makebox[0pt][r]{\rotatebox[origin=rB]{45}{\textbf{Rapid LLM}}}} & \multicolumn{1}{c}{\makebox[0pt][r]{\rotatebox[origin=rB]{45}{\textbf{Rapid Agent}}}} & \multicolumn{1}{c}{\makebox[0pt][r]{\rotatebox[origin=rB]{45}{\textbf{Gradual AI}}}} & \multicolumn{1}{c}{\makebox[0pt][r]{\rotatebox[origin=rB]{45}{\textbf{Rapid LLM}}}} & \multicolumn{1}{c}{\makebox[0pt][r]{\rotatebox[origin=rB]{45}{\textbf{Rapid Agent}}}} & \multicolumn{1}{c}{\makebox[0pt][r]{\rotatebox[origin=rB]{45}{\textbf{Gradual AI}}}} & \multicolumn{1}{c}{\makebox[0pt][r]{\rotatebox[origin=rB]{45}{\textbf{Rapid LLM}}}} & \multicolumn{1}{c}{\makebox[0pt][r]{\rotatebox[origin=rB]{45}{\textbf{Rapid Agent}}}} & \multicolumn{1}{c}{\makebox[0pt][r]{\rotatebox[origin=rB]{45}{\textbf{Gradual AI}}}} & \multicolumn{1}{c}{\makebox[0pt][r]{\rotatebox[origin=rB]{45}{\textbf{Rapid LLM}}}} & \multicolumn{1}{c}{\makebox[0pt][r]{\rotatebox[origin=rB]{45}{\textbf{Rapid Agent}}}} & \multicolumn{1}{c}{\makebox[0pt][r]{\rotatebox[origin=rB]{45}{\textbf{Gradual AI}}}} & \multicolumn{1}{c}{\makebox[0pt][r]{\rotatebox[origin=rB]{45}{\textbf{Rapid LLM}}}} & \multicolumn{1}{c}{\makebox[0pt][r]{\rotatebox[origin=rB]{45}{\textbf{Rapid Agent}}}} & \multicolumn{1}{c}{\makebox[0pt][r]{\rotatebox[origin=rB]{45}{\textbf{Gradual AI}}}} & \multicolumn{1}{c}{\makebox[0pt][r]{\rotatebox[origin=rB]{45}{\textbf{Rapid LLM}}}} & \multicolumn{1}{c}{\makebox[0pt][r]{\rotatebox[origin=rB]{45}{\textbf{Rapid Agent}}}} & \multicolumn{1}{c}{\makebox[0pt][r]{\rotatebox[origin=rB]{45}{\textbf{Gradual AI}}}} & \multicolumn{1}{c}{\makebox[0pt][r]{\rotatebox[origin=rB]{45}{\textbf{Rapid LLM}}}} & \multicolumn{1}{c}{\makebox[0pt][r]{\rotatebox[origin=rB]{45}{\textbf{Rapid Agent}}}} & \multicolumn{1}{c}{\makebox[0pt][r]{\rotatebox[origin=rB]{45}{\textbf{Gradual AI}}}} & \multicolumn{1}{c}{\makebox[0pt][r]{\rotatebox[origin=rB]{45}{\textbf{Rapid LLM}}}} & \multicolumn{1}{c}{\makebox[0pt][r]{\rotatebox[origin=rB]{45}{\textbf{Rapid Agent}}}} & \multicolumn{1}{c}{\makebox[0pt][r]{\rotatebox[origin=rB]{45}{\textbf{Gradual AI}}}} & \multicolumn{1}{c}{\makebox[0pt][r]{\rotatebox[origin=rB]{45}{\textbf{Rapid LLM}}}} & \multicolumn{1}{c}{\makebox[0pt][r]{\rotatebox[origin=rB]{45}{\textbf{Rapid Agent}}}} & \multicolumn{1}{c}{\makebox[0pt][r]{\rotatebox[origin=rB]{45}{\textbf{Gradual AI}}}} & \multicolumn{1}{c}{\makebox[0pt][r]{\rotatebox[origin=rB]{45}{\textbf{Rapid LLM}}}} & \multicolumn{1}{c}{\makebox[0pt][r]{\rotatebox[origin=rB]{45}{\textbf{Rapid Agent}}}} & \multicolumn{1}{c}{\makebox[0pt][r]{\rotatebox[origin=rB]{45}{\textbf{Gradual AI}}}} & \multicolumn{1}{c}{\makebox[0pt][r]{\rotatebox[origin=rB]{45}{\textbf{Rapid LLM}}}} & \multicolumn{1}{c}{\makebox[0pt][r]{\rotatebox[origin=rB]{45}{\textbf{Rapid Agent}}}} & \multicolumn{1}{c}{\makebox[0pt][r]{\rotatebox[origin=rB]{45}{\textbf{Gradual AI}}}} & \multicolumn{1}{c}{\makebox[0pt][r]{\rotatebox[origin=rB]{45}{\textbf{Rapid LLM}}}} & \multicolumn{1}{c}{\makebox[0pt][r]{\rotatebox[origin=rB]{45}{\textbf{Rapid Agent}}}} \\
\end{tabular}
}
\begin{center}
\end{center}
\vspace{-30pt}
\end{table*}

\textbf{Estimating Factor Importance.} We follow prior modeling analyses~\cite{tourani2016impact, thongtanunam2017review} by adding our explanatory features to the model one at a time and applying a likelihood ratio chi-square test~\cite{Hahn01101984}. A $p$-value below 0.05 from the likelihood ratio chi-square test indicates that the newly added factor significantly improves model fit beyond the factors already included. For example, consider a logistic regression model with efficient review as the dependent variable, where reviewer collaboration pattern and pull request type are already included as independent variables. When review activity is added as another independent variable, the likelihood-ratio chi-square test compares the expanded model with the previous model that includes only reviewer collaboration pattern and pull request type. A $p$-value below 0.05 shows that review activity significantly improves model fit after those variables have already been considered. This analysis helps determine which variables (e.g., review activity) have an independent explanatory contribution to review efficiency or review smells. To this end, we first measure a baseline prediction~(\(P_{\text{base}}\)) by setting all variables to their baseline values~(i.e., median values for numeric variables and reference categories for categorical variables). We then compute a second prediction~(\(P_{\text{changed}}\)) by changing only the target variable by a typical amount~(i.e., increasing a numeric variable by one standard deviation or setting a categorical variable to the target category), while keeping all other variables at their baseline values~\cite{shihab2013lines, thongtanunam2017review}. We then calculate the impact percentage as follows:
\vspace{-2pt}
\begin{equation}
\text{Impact Score~(Imp\%)} = \frac{P_{\text{changed}} - P_{\text{base}}}{P_{\text{base}}} \times 100\%
\vspace{-2pt}
\end{equation}

Positive Imp\% indicates a higher predicted probability of the modeled outcome, and negative Imp\% indicates a lower predicted probability. By dividing each effect by the baseline probability~($P_{\text{base}}$), Imp\% places variables with different units and ranges on a common scale. We can therefore directly compare the practical impact of each significant variable~(e.g., the number of reviewers) and identify which factors are most strongly associated with review efficiency or review smells.

\findings 
We organize our findings by review efficiency and three selected review smells. We discuss each outcome below.

\noindent\textit{(1) Review Efficiency.}~\textbf{Greater reliance on AI reviewers does not necessarily translate into more efficient code reviews, as traditional review factors continue to play an important role.} Table~\ref{tab:rq3_effi_smell_byera} presents the impact scores of the significant explanatory factors for each outcome. In the Pre-LLM era, author experience and pull request characteristics (e.g., chore and refactoring pull request types) are the primary factors associated with review efficiency. For example, author experience is associated with lower review delay across the three adoption practices, with impact scores from -17 to -22 in the Pre-LLM era. In the LLM era, collaboration patterns involving LLMs or bots are generally associated with higher review delay (i.e., lower efficiency), while traditional review factors remain relevant but less consistently, particularly for author experience and refactoring. In the agent era, AI agent collaboration becomes the dominant factor associated with review efficiency. Specifically, multi-agent and agent-init collaboration patterns are negatively associated with review delay~(i.e., higher efficiency) under Gradual AI Adoption and Rapid AI Agent Adoption. However, under Rapid LLM Adoption, where traditional review factors are no longer significant, AI collaboration patterns remain consistently associated with higher delay and therefore lower review efficiency, suggesting that projects relying more on AI-assisted reviews, without the influence of traditional review factors, do not necessarily achieve more efficient reviews.

\noindent\textit{(2) Review Buddies.}~\textbf{Build pull requests are often associated with fewer Review Buddies across review eras, suggesting more diverse reviewer participation for maintenance-oriented tasks.} As shown in Table~\ref{tab:rq3_effi_smell_byera}, build, chore, and refactoring pull requests exhibit negative associations with Review Buddies under the Rapid LLM Adoption in the LLM era ($-43$, $-39$, and $-19$) and the Agent era ($-55$, $-25$, and $-12$). This pattern is also observed for build pull requests under Rapid AI Agent Adoption, where build pull requests exhibit negative associations with Review Buddies across the three eras ($-15$, $-20$, and $-19$). This suggests that these maintenance-oriented pull requests are less likely to rely on repeated reviewer pairings and instead involve a more diverse set of reviewers. However, refactoring pull requests show positive associations with Review Buddies under Rapid AI Agent Adoption across the three eras ($+29$, $+20$, and $+33$), indicating more repeated reviewer pairings for this pull request type. Furthermore, author experience is consistently positively associated with Review Buddies across all review eras, indicating that more experienced authors are more likely to review with recurring collaborators, whereas less experienced authors tend to participate in more diverse review interactions.

\noindent\textit{(3) Sleeping Review.}~\textbf{In the Agent era, multi-agent and agent-init reviews are associated with fewer Sleeping Reviews under Gradual AI Adoption and Rapid AI Agent Adoption.} Across the three review eras, pull requests with more commits and more inline discussion threads are consistently associated with a higher probability of Sleeping Review, with impact scores ranging from $+27$ to $+70$ for commits and from $+26$ to $+51$ for inline discussion threads. By contrast, pull requests submitted by authors with more prior experience in the same project, as well as build and chore pull requests, are associated with a lower probability of Sleeping Review, with impact scores ranging from $-10$ to $-50$. In the LLM era, LLM-Assist and LLM-Bot-Assist are associated with a higher probability of Sleeping Review, with impact scores ranging from $+26$ to $+214$. In the agent era, Multi-Agent and Agent-Init patterns are associated with a lower probability of Sleeping Review in Gradual AI Adoption and Rapid AI Agent Adoption, with impact scores ranging from $-16$ to $-48$. By contrast, under Rapid LLM Adoption, agent patterns such as Agent-Init and Agent-ML are associated with a much higher probability of Sleeping Review, with impact scores of $+42$ and $+94$.

\noindent\textit{(4) Large Changeset.}~\textbf{Agent-involved collaboration patterns under Gradual AI Adoption and Rapid AI Agent Adoption are often associated with a higher likelihood of Large Changeset.} As shown in Table~\ref{tab:rq3_effi_smell_byera}, among traditional factors, only build pull requests show mixed associations with Large Changeset across practices and eras. Other factors, including commit count, inline discussion threads, author experience, chore and refactoring pull requests, show positive significant associations where they appear. In the agent era, significant agent-involved collaboration patterns show positive associations with Large Changeset under Gradual AI Adoption and Rapid AI Agent Adoption, with impact scores ranging from $+25$ to $+48$ and from $+20$ to $+54$, respectively. These findings suggest that agent-involved collaboration is more frequently associated with the review of larger code changes than in the Pre-LLM and LLM eras.

\smallskip
\vspace{-5pt}
\begin{Summary}{Summary for RQ3}{}
Traditional review factors remain important after generative AI reviewers are adopted. Agent-era collaboration patterns reduce Sleeping Review under Gradual AI Adoption and Rapid AI Agent Adoption, but they are also associated with more Review Buddies and larger changesets.
\end{Summary}
\vspace{-5pt}

\section{Implications}\label{sec:implications}
In this section, we discuss the implications of our findings for practitioners, tool builders, and researchers.

\textbf{Adopting AI reviewers selectively rather than uniformly.}
Our results show that AI reviewer adoption practices are associated with differences in review quality. Gradual AI Adoption and Rapid AI Agent Adoption are associated with faster reviews, while Rapid LLM Adoption is associated with lower review quality. Therefore, practitioners may consider adopting AI reviewers selectively, based on pull request context and observed review efficiency and quality, rather than applying the same AI reviewer configuration uniformly across pull requests.

\textbf{Building context-aware support for human-AI review decisions.}
Our findings identify several signals that can inform AI reviewer assignment, including human-AI collaboration patterns, pull request type, changeset size, discussion activity, and author or reviewer history. Tool builders can use these signals to help practitioners decide whether a pull request should remain human-led, use lightweight LLM summarization, or involve AI agent inspection.

\textbf{Balancing AI collaboration with traditional review practices.}
Our findings show that AI collaboration does not uniformly improve review quality. While some agent-era collaboration patterns are associated with fewer Sleeping Reviews, they are also associated with more Review Buddies and larger changesets. Traditional review factors, such as author experience and review activity, remain strongly associated with review efficiency and review smells. Practitioners should therefore integrate AI reviewers alongside established review practices rather than relying on AI collaboration alone.

\textbf{Studying developer decisions behind human-AI review practices.}
AI review practices vary across projects, but our results do not explain why developers choose particular human-AI review practices. Future research can build on our results through interviews or surveys to understand why developers assign AI reviewers to some pull requests, how they interpret AI feedback, when they keep humans in the lead, and how these decisions shape review efficiency and quality.

\textbf{Developing benchmarks that reflect pull request context.}
Evaluating AI reviewers only by generic issue-finding ability may miss important differences across review contexts. Researchers and tool builders can use signals such as pull request type, changeset size, discussion activity, author or reviewer history, and human-AI collaboration pattern to design benchmarks across contexts such as routine maintenance, bug fixing, refactoring, and large changesets. Such benchmarks would help assess when AI reviewers provide useful support and when human-led review remains necessary.
\section{Threats to Validity}\label{sec:threatstovalidity}

\textbf{Threats to construct validity.} 
In our analysis, the agent era denotes a chronological period in a project's review history rather than guaranteeing that every AI reviewer during that period is an AI agent. A project enters the agent era after its first pull request involving an AI agent reviewer, but may continue to use other LLM reviewers afterward. Therefore, agent-era pull requests can still contain LLM reviewer participation. For example, the identified collaboration patterns include LLM-involved patterns in the agent era, such as LLM-Assist. However, these older patterns do not have sufficient support~(less than 5\%), so we exclude them from the pairwise comparisons reported in Table~\ref{tab:rq2_patterns_v2}. The complete pattern results and their frequencies are available in our replication package~\cite{github_ai_review_evolution}.

\textbf{Threats to internal validity.} Our logistic regression models in RQ3 are explanatory, not causal. The associations we report reflect statistical relationships between factors and review efficiency and quality within each practice and era, but do not imply that changing one factor will directly cause a change in review efficiency or review smells.

\textbf{Threats to external validity.} Given the rapid evolution of AI reviewer capabilities, the collaboration patterns and review quality associations that we observe may shift as tools become more sophisticated. Our dataset covers activity from May 2022 through February 2026, providing a current longitudinal view of AI reviewer adoption. While findings specific to individual tools may evolve, our longitudinal analysis across the pre-LLM, LLM, and agent eras provides a baseline and reference point for future studies to track how AI reviewer adoption and review quality continue to change.

\section{Related work}\label{sec:relatedwork}
\textbf{Code review practices.}
Code review is a fundamental quality assurance practice where developers examine code changes before integration. Prior work has shown how review intervals and participation converge across diverse organizations~\cite{rigby2013convergent}, characterized modern review practices at scale~\cite{sadowski2018modern}, introduced code review smell taxonomies~\cite{reviewsmell2022}, and shown that reviewer experience and changeset characteristics influence review comment quality~\cite{bosu2015characteristics}. 
As AI reviewers enter code review, recent studies have evaluated LLM-assisted review tools in specific settings. Cihan et al.~\cite{cihan2025automatedcodereview} find that LLM-assisted review may increase pull request review time despite high comment resolution rates. A{\dh}alsteinsson et al.~\cite{adhalsteinsson2025rethinking} find that LLM-assisted review reduces reviewer cognitive load and improves comprehension. Sun et al.~\cite{sun2025doesaicodereview} find that LLM reviewers produce concise comments that are more likely to lead to code changes. 
In this study, we move beyond evaluating individual AI review tools in isolation and analyze how review practices evolve as projects transition from human-centric review to agentic review. We examine this transition at the project level and pull-request level by identifying AI reviewer adoption practices and human-AI collaboration patterns.

\textbf{Review efficiency and quality factors.}
Prior work has studied technical and social factors associated with code review efficiency and quality in human-centric review processes. The factors include organizational structure and developer relationships~\cite{baysal2016investigating}, past review participation and patch description length~\cite{thongtanunam2017review}, and reviewer experience and changeset characteristics~\cite{bosu2015characteristics}. Such factors help explain why some reviews complete faster or produce more useful feedback, but they were developed mainly for human-only review settings. In this study, we extend these review outcome factors to the transition from human-centric to agentic review. We model review outcomes together with human-AI collaboration patterns, pull request characteristics, review activity, and participant experience to examine whether efficiency and quality differences are associated mainly with AI reviewer participation or also with the alternative factors.

\section{Conclusion}\label{sec:conclusion}

In this study, we analyze 1.02 million pull requests from 207 GitHub projects to understand how code review quality evolves across three review eras. We identify three AI reviewer adoption practices and find that Gradual AI Adoption and Rapid AI Agent Adoption are associated with faster reviews but no improvement in review smells, while Rapid LLM Adoption is associated with higher review-smell prevalence and no efficiency gain. Within individual pull requests, agent-init and multi-agent reviews reach a decision faster under Gradual AI and Rapid AI Agent Adoption, while they carry review smells more often than human-only reviews. We further show that, beyond traditional factors, human-AI collaboration patterns remain strongly associated with review efficiency and quality. 

Together, these findings suggest that developers should adopt AI reviewers selectively, taking into account their adoption history, who reviews each pull request and when, and the characteristics of each pull request. In future work, we aim to study how agentic review systems can improve review efficiency while preserving the quality standards established in human code review.



%
%


\balance
\bibliographystyle{IEEEtran}

\bibliography{Bib}

@article{watanabe2025useagenticcodingempirical,
author = {Watanabe, Miku and Li, Hao and Kashiwa, Yutaro and Reid, Brittany and Iida, Hajimu and Hassan, Ahmed E.},
title = {On the Use of Agentic Coding: An Empirical Study of Pull Requests on GitHub},
year = {2026},
publisher = {Association for Computing Machinery},
address = {New York, NY, USA},
issn = {1049-331X},
url = {https://doi.org/10.1145/3798166},
doi = {10.1145/3798166},
note = {Just Accepted},
journal = {ACM Transactions on Software Engineering and Methodology},
month = mar
}

@misc{israel1992determining,
  title = {A systematic review of statistical power in software engineering experiments},
journal = {Information and Software Technology},
volume = {48},
number = {8},
pages = {745-755},
year = {2006},
issn = {0950-5849},
doi = {https://doi.org/10.1016/j.infsof.2005.08.009},
url = {https://doi.org/10.1016/j.infsof.2005.08.009
},
author = {Tore Dybå and Vigdis By Kampenes and Dag I.K. Sjøberg},
}

@Inbook{Hahn01101984,
author = {Boslaugh, Sarah and Watters, Dr. Paul A.},
title = {Statistics in a nutshell},
year = {2008},
isbn = {9780596510497},
publisher = {O'Reilly \& Associates, Inc.},
address = {USA},
edition = {First},
url={https://dl.acm.org/doi/book/10.5555/1461408}
}

@misc{wikipedia_chatgpt,
  author = {{Wikipedia contributors}},
  title = {{ChatGPT}},
  howpublished = {\url{https://en.wikipedia.org/wiki/ChatGPT}},
  year = {2025},
  note = {Accessed: 2025-12-30}
}

@misc{github_ai_review_evolution,
  author = {{Anonymous}},
  title = {{AI} Review Evolution},
  howpublished = {\url{https://anonymous.4open.science/r/CodeReviewEvolve-7917}},
  year = {2026},
  note = {Accessed: 2026-02-26}
}

@ARTICLE{sun2025doesaicodereview,
   author={Sun, Kexin and Kuang, Hongyu and Baltes, Sebastian and Zhou, Xin and Zhang, He and Ma, Xiaoxing and Rong, Guoping and Shao, Dong and Treude, Christoph},
  journal={IEEE Transactions on Software Engineering}, 
  title={Does AI Code Review Lead to Code Changes? A Case Study of GitHub Actions}, 
  year={2026},
  volume={},
  number={},
  pages={1-17},
  doi={10.1109/TSE.2026.3688237},
  url={https://doi.org/10.1109/TSE.2026.3688237}}

@misc{rasheed2024aipoweredcodereviewllms,
      title={AI-powered Code Review with LLMs: Early Results}, 
      author={Zeeshan Rasheed and Malik Abdul Sami and Muhammad Waseem and Kai-Kristian Kemell and Xiaofeng Wang and Anh Nguyen and Kari Systä and Pekka Abrahamsson},
      year={2025},
      eprint={2404.18496},
      archivePrefix={arXiv},
      primaryClass={cs.SE},
      url={https://arxiv.org/abs/2404.18496}, 
}

@article{friedman1962interpolation,
  AUTHOR = {Lepot, Mathieu and Aubin, Jean-Baptiste and Clemens, François H.L.R.},
TITLE = {Interpolation in Time Series: An Introductive Overview of Existing Methods, Their Performance Criteria and Uncertainty Assessment},
JOURNAL = {Water},
VOLUME = {9},
YEAR = {2017},
NUMBER = {10},
ARTICLE-NUMBER = {796},
URL = {https://doi.org/10.3390/w9100796},
ISSN = {2073-4441},
DOI = {10.3390/w9100796}
}

@article{rousseeuw1987silhouettes,
  title = {Silhouettes: A graphical aid to the interpretation and validation of cluster analysis},
journal = {Journal of Computational and Applied Mathematics},
volume = {20},
pages = {53-65},
year = {1987},
issn = {0377-0427},
doi = {https://doi.org/10.1016/0377-0427(87)90125-7},
url = {https://doi.org/10.1016/0377-0427(87)90125-7},
author = {Peter J. Rousseeuw}
}

@misc{li2025riseaiteammatessoftware,
      title={The Rise of AI Teammates in Software Engineering (SE) 3.0: How Autonomous Coding Agents Are Reshaping Software Engineering}, 
      author={Hao Li and Haoxiang Zhang and Ahmed E. Hassan},
      year={2025},
      eprint={2507.15003},
      archivePrefix={arXiv},
      primaryClass={cs.SE},
      url={https://arxiv.org/abs/2507.15003}, 
}

@article{cohen1960coefficient,
  author={Emam, Khaled El},
title={Benchmarking Kappa: Interrater Agreement in Software Process Assessments},
journal={Empirical Software Engineering},
year={1999},
month={Jun},
day={01},
volume={4},
number={2},
pages={113-133},
issn={1573-7616},
doi={10.1023/A:1009820201126},
url={https://doi.org/10.1023/A:1009820201126}
}

@article{mcintosh2016empirical,
  author={McIntosh, Shane
and Kamei, Yasutaka
and Adams, Bram
and Hassan, Ahmed E.},
title={An empirical study of the impact of modern code review practices on software quality},
journal={Empirical Software Engineering},
year={2016},
month={Oct},
day={01},
volume={21},
number={5},
pages={2146-2189},
issn={1573-7616},
doi={10.1007/s10664-015-9381-9},
url={https://doi.org/10.1007/s10664-015-9381-9}
}

@inproceedings{unmaintain_project_github,
author = {Coelho, Jailton and Valente, Marco Tulio and Silva, Luciana L. and Shihab, Emad},
title = {Identifying unmaintained projects in github},
year = {2018},
isbn = {9781450358231},
publisher = {Association for Computing Machinery},
address = {New York, NY, USA},
url = {https://doi.org/10.1145/3239235.3240501},
doi = {10.1145/3239235.3240501},
booktitle = {Proceedings of the 12th ACM/IEEE International Symposium on Empirical Software Engineering and Measurement},
articleno = {15},
numpages = {10},
location = {Oulu, Finland},
series = {ESEM '18}
}

@misc{github_rest_api,
  author = {{GitHub}},
  title = {{GitHub} {REST} {API}},
  howpublished = {\url{https://docs.github.com/en/rest?apiVersion=2026-03-10
}},
  year = {2026},
  month = Feb,
  note = {Accessed: 2026-2-24}
}

@misc{github_search,
  author = {{GitHub}},
  title = {{GitHub} Advanced Search},
  howpublished = {\url{https://github.com/search/advanced}},
  year = {2026},
  month = Feb,
  note = {Accessed: 2026-06-30}
}

@misc{CopilotAgent2025,
  author = {{Thomas Dohmke}},
  title = {Coding Agent for {Copilot}},
  howpublished = {\url{https://github.blog/news-insights/product-news/github-copilot-meet-the-new-coding-agent/}},
  year = {2025},
  month = may,
  day = {19},
  note = {Accessed: 2026-01-01}
}

@misc{GoogleReviewBottleneck2026,
  author = {{Lee Boonstra}},
  title = {Google Review Bottleneck},
  howpublished = {\url{https://cloud.google.com/transform/when-ai-writes-the-code-who-reviews-it-cto-google-cloud}},
  year = {2026},
  month = April,
  day = {28},
  note = {Accessed: 2026-06-16}
}

@misc{AmazonReviewBottleneck2026,
  author = {{Matthias Patzak}},
  title = {Amazon Review Bottleneck},
  howpublished ={\url{https://aws.amazon.com/blogs/enterprise-strategy/your-ai-coding-assistants-will-overwhelm-your-delivery-pipeline-heres-how-to-prepare}},
  year = {2026},
  month = Jan,
  day = {07},
  note = {Accessed: 2026-06-16}
}

@misc{ClaudeCode2025,
  author = {{Anthropic}},
  title = {{Claude 3.7 Sonnet and Claude Code}},
  howpublished = {\url{https://www.anthropic.com/news/claude-3-7-sonnet}},
  year = {2025},
  month = feb,
  day = {24},
  note = {Accessed: 2026-05-25}
}

@misc{GithubActions,
  author = {{GitHub, Inc.}},
  title = {{GitHub} Actions Bot Details},
  howpublished = {\url{https://github.com/marketplace/actions/bot-details}},
  year = {2026},
  month = Jan,
  day = {19},
  note = {Accessed: 2026-01-01}
}

@inproceedings{bose2009context,
  author = {R. P. Jagadeesh Chandra Bose and Wil M.P. van der Aalst},
title = {Context Aware Trace Clustering: Towards Improving Process Mining Results},
booktitle = {Proceedings of the 2009 SIAM International Conference on Data Mining (SDM)},
chapter = {},
pages = {401-412},
doi = {10.1137/1.9781611972795.35},
url = {https://doi.org/10.1137/1.9781611972795.35},
eprint = {https://epubs.siam.org/doi/pdf/10.1137/1.9781611972795.35},
}

@inproceedings{song2008trace,
  author="Song, Minseok
and G{\"u}nther, Christian W.
and van der Aalst, Wil M. P.",
editor="Ardagna, Danilo
and Mecella, Massimo
and Yang, Jian",
title="Trace Clustering in Process Mining",
booktitle="Business Process Management Workshops",
year="2009",
publisher="Springer Berlin Heidelberg",
address="Berlin, Heidelberg",
pages="109--120",
isbn="978-3-642-00328-8",
url={https://doi.org/10.1007/978-3-642-00328-8_11}
}

@article{moon1996expectation,
   author={Moon, T.K.},
  journal={IEEE Signal Processing Magazine}, 
  title={The expectation-maximization algorithm}, 
  year={1996},
  volume={13},
  number={6},
  pages={47-60},
  doi={10.1109/79.543975},
  url={https://doi.org/10.1109/79.543975}}

@article{watanabe2013widely,
  author = {Watanabe, Sumio},
title = {A widely applicable Bayesian information criterion},
year = {2013},
issue_date = {January 2013},
publisher = {JMLR.org},
volume = {14},
number = {1},
issn = {1532-4435},
journal = {The Journal of Machine Learning Research},
month = mar,
pages = {867–897},
numpages = {31},
url={https://dl.acm.org/doi/10.5555/2567709.2502609}
}

@article{jelihovschi2014scottknott,
  title={ScottKnott: A Package for Performing the Scott-Knott Clustering Algorithm in R}, 
  volume={15}, 
  url={https://doi.org/10.5540/tema.2014.015.01.0003}, 
  DOI={10.5540/tema.2014.015.01.0003}, number={1}, 
  journal={Trends in Computational and Applied Mathematics}, 
  author={Jelihovschi, Enio and Faria, José Cláudio and Allaman, Ivan Bezerra}, 
  year={2014}, 
  month={Mar.}, 
  pages={003–017} 
  }

@inproceedings{izquierdo2017using,
  author = {Izquierdo-Cortazar, Daniel and Sekitoleko, Nelson and Gonzalez-Barahona, Jesus M. and Kurth, Lars},
title = {Using Metrics to Track Code Review Performance},
year = {2017},
isbn = {9781450348041},
publisher = {Association for Computing Machinery},
address = {New York, NY, USA},
url = {https://doi.org/10.1145/3084226.3084247},
doi = {10.1145/3084226.3084247},
booktitle = {Proceedings of the 21st International Conference on Evaluation and Assessment in Software Engineering},
pages = {214–223},
numpages = {10},
location = {Karlskrona, Sweden},
series = {EASE '17}
}

@inproceedings{tourani2016impact,
  author={Tourani, Parastou and Adams, Bram},
  booktitle={2016 IEEE 23rd International Conference on Software Analysis, Evolution, and Reengineering (SANER)}, 
  title={The Impact of Human Discussions on Just-in-Time Quality Assurance: An Empirical Study on OpenStack and Eclipse}, 
  year={2016},
  number={},
  pages={189-200},
  doi={10.1109/SANER.2016.113},
  url={https://doi.org/10.1109/SANER.2016.113}
  }

@article{ghaleb2019studying,
   author={Ghaleb, Taher Ahmed and da Costa, Daniel Alencar and Zou, Ying and Hassan, Ahmed E.},
  journal={IEEE Transactions on Software Engineering}, 
  title={Studying the Impact of Noises in Build Breakage Data}, 
  year={2021},
  volume={47},
  number={9},
  pages={1998-2011},
  doi={10.1109/TSE.2019.2941880},
  url={https://doi.org/10.1109/TSE.2019.2941880}}

@article{bradley1997use,
  title = {The use of the area under the ROC curve in the evaluation of machine learning algorithms},
journal = {Pattern Recognition},
volume = {30},
number = {7},
pages = {1145-1159},
year = {1997},
issn = {0031-3203},
doi = {https://doi.org/10.1016/S0031-3203(96)00142-2},
url = {https://doi.org/10.1016/S0031-3203(96)00142-2},
author = {Andrew P. Bradley}
}

@article{thongtanunam2017review,
  author={Thongtanunam, Patanamon
and McIntosh, Shane
and Hassan, Ahmed E.
and Iida, Hajimu},
title={Review participation in modern code review},
journal={Empirical Software Engineering},
year={2017},
month={Apr},
day={01},
volume={22},
number={2},
pages={768-817},
issn={1573-7616},
doi={10.1007/s10664-016-9452-6},
url={https://doi.org/10.1007/s10664-016-9452-6}
}

@inproceedings{rigby2013convergent,
  author = {Rigby, Peter C. and Bird, Christian},
title = {Convergent contemporary software peer review practices},
year = {2013},
isbn = {9781450322379},
publisher = {Association for Computing Machinery},
address = {New York, NY, USA},
url = {https://doi.org/10.1145/2491411.2491444},
doi = {10.1145/2491411.2491444},
booktitle = {Proceedings of the 2013 9th Joint Meeting on Foundations of Software Engineering},
pages = {202–212},
numpages = {11},
location = {Saint Petersburg, Russia},
series = {ESEC/FSE 2013}
}

@article{baysal2016investigating,
  author={Baysal, Olga
and Kononenko, Oleksii
and Holmes, Reid
and Godfrey, Michael W.},
title={Investigating technical and non-technical factors influencing modern code review},
journal={Empirical Software Engineering},
year={2016},
month={Jun},
day={01},
volume={21},
number={3},
pages={932-959},
issn={1573-7616},
doi={10.1007/s10664-015-9366-8},
url={https://doi.org/10.1007/s10664-015-9366-8}
}

@inproceedings{sadowski2018modern,
  author = {Sadowski, Caitlin and S\"{o}derberg, Emma and Church, Luke and Sipko, Michal and Bacchelli, Alberto},
title = {Modern code review: a case study at google},
year = {2018},
isbn = {9781450356596},
publisher = {Association for Computing Machinery},
address = {New York, NY, USA},
url = {https://doi.org/10.1145/3183519.3183525},
doi = {10.1145/3183519.3183525},
booktitle = {Proceedings of the 40th International Conference on Software Engineering: Software Engineering in Practice},
pages = {181–190},
numpages = {10},
location = {Gothenburg, Sweden},
series = {ICSE-SEIP '18}
}

@inproceedings{bosu2015characteristics,
  author={Bosu, Amiangshu and Greiler, Michaela and Bird, Christian},
  booktitle={2015 IEEE/ACM 12th Working Conference on Mining Software Repositories}, 
  title={Characteristics of Useful Code Reviews: An Empirical Study at Microsoft}, 
  year={2015},
  volume={},
  number={},
  pages={146-156},
  doi={10.1109/MSR.2015.21},
  url={https://doi.org/10.1109/MSR.2015.21}}

@article{shihab2013lines,
  title = {Is lines of code a good measure of effort in effort-aware models?},
journal = {Information and Software Technology},
volume = {55},
number = {11},
pages = {1981-1993},
year = {2013},
issn = {0950-5849},
doi = {https://doi.org/10.1016/j.infsof.2013.06.002},
url = {https://doi.org/10.1016/j.infsof.2013.06.002},
author = {Emad Shihab and Yasutaka Kamei and Bram Adams and Ahmed E. Hassan},
}

@article{reviewsmell2022,
title = {Towards a taxonomy of code review smells},
journal = {Information and Software Technology},
volume = {142},
pages = {106737},
year = {2022},
issn = {0950-5849},
doi = {https://doi.org/10.1016/j.infsof.2021.106737},
url = {https://doi.org/10.1016/j.infsof.2021.106737
},
author = {Emre Doğan and Eray Tüzün},
}

@misc{AWSCodeGuru,
  author = {{Amazon Web Services}},
  title = {{Amazon CodeGuru Reviewer}},
  howpublished = {\url{https://docs.aws.amazon.com/codeguru/latest/reviewer-ug/welcome.html}},
  year = {2025},
  note = {Accessed: 2026-04-09}
}

@INPROCEEDINGS{cihan2025automatedcodereview,
  author={Cihan, Umut and Haratian, Vahid and {\.I}\c{c}{\"{o}}z, Arda and G{\"{u}}l, Mert Kaan and Devran, {\"O}mercan and Bayendur, Emircan Furkan and U\c{c}ar, Baykal Mehmet and T{\"{u}}z{\"{u}}n, Eray},
    booktitle={2025 IEEE/ACM 47th International Conference on Software Engineering: Software Engineering in Practice (ICSE-SEIP)}, 
  title={Automated Code Review in Practice}, 
  year={2025},
  volume={},
  number={},
  pages={425-436},
  doi={10.1109/ICSE-SEIP66354.2025.00043},
  url={https://doi.org/10.1109/ICSE-SEIP66354.2025.00043}}

@misc{zhong2026humanaisynergyagenticcode,
      title={Human-AI Synergy in Agentic Code Review}, 
      author={Suzhen Zhong and Shayan Noei and Ying Zou and Bram Adams},
      year={2026},
      eprint={2603.15911},
      archivePrefix={arXiv},
      primaryClass={cs.SE},
      url={https://arxiv.org/abs/2603.15911}, 
}

@inproceedings{adhalsteinsson2025rethinking,
  title={Rethinking code review workflows with llm assistance: An empirical study},
  author={A{\dh}alsteinsson, Fannar Steinn and Magn{\'u}sson, Bj{\"o}rn Borgar and Milicevic, Mislav and Davidsson, Adam Nirving and Cheng, Chih-Hong},
    booktitle={2025 ACM/IEEE International Symposium on Empirical Software Engineering and Measurement (ESEM)}, 
  year={2025},
  volume={},
  number={},
  pages={488-497},
  doi={10.1109/ESEM64174.2025.00013},
  url={https://doi.org/10.1109/ESEM64174.2025.00013}}

@article{noei2025detecting,
  author = {Noei, Shayan and Li, Heng and Zou, Ying},
title = {Detecting Refactoring Commits in Machine Learning Python Projects: A Machine Learning-Based Approach},
year = {2025},
issue_date = {March 2025},
publisher = {Association for Computing Machinery},
address = {New York, NY, USA},
volume = {34},
number = {3},
issn = {1049-331X},
url = {https://doi.org/10.1145/3705309},
doi = {10.1145/3705309},
journal = {ACM Transactions on Software Engineering and Methodology},
month = feb,
articleno = {68},
numpages = {25}
}

@article{yu2026empirical,
  author = {Yu, Chunli and Noei, Shayan and Zhang, Haoxiang and Zou, Ying},
title = {An Empirical Study on the Characteristics of Reusable Code Clones},
year = {2026},
publisher = {Association for Computing Machinery},
address = {New York, NY, USA},
issn = {1049-331X},
url = {https://doi.org/10.1145/3793251},
doi = {10.1145/3793251},
note = {Just Accepted},
journal = {ACM Transactions on Software Engineering and Methodology},
month = jan
}

@article{ouf2026empirical,
  title={An Empirical Analysis of Community and Coding Patterns in OSS4SG vs. Conventional OSS},
  author={Ouf, Mohamed and Noei, Shayan and Van Iterson, Zeph and Guizani, Mariam and Zou, Ying},
  journal={arXiv preprint arXiv:2601.03430},
  year={2026}
}

@misc{google_ai_code_75,
  title={Google says 75\% of the new code is {AI}-generated},
  author={{Hugh Langley}},
  year={2026},
  month={April},
  howpublished={\url{https://www.businessinsider.com/google-ai-generated-code-75-gemini-agents-software-2026-4}},
  note={Accessed: 2026-06-30}
}

@misc{github_octoverse_2025,
  title={A new {AI} developer joins {GitHub}},
  author={{GitHub}},
  year={2025},
  month={October},
  howpublished={\href{https://github.blog/news-insights/octoverse/octoverse-a-new-developer-joins-github-every-second-as-ai-leads-typescript-to-1/}{https://github.blog/news-insights/octoverse/octoverse-a-new-developer-joins-github-every-second-as-ai-leads-typescript-to-1/}},
  note={Updated February 28, 2026. Accessed: 2026-06-30}
}

@article{noei2025empirical,
  author = {Noei, Shayan and Li, Heng and Zou, Ying},
title = {An Empirical Study on Release-Wise Refactoring Patterns},
year = {2025},
issue_date = {July 2025},
publisher = {Association for Computing Machinery},
address = {New York, NY, USA},
volume = {2},
number = {FSE},
url = {https://doi.org/10.1145/3715734},
doi = {10.1145/3715734},
journal = {Proceedings of the ACM on Software Engineering},
month = jun,
articleno = {FSE019},
numpages = {22}
}

@INPROCEEDINGS{ReviewBots2020,
    author={Wessel, Mairieli and Serebrenik, Alexander and Wiese, Igor and Steinmacher, Igor and Gerosa, Marco A.},
  booktitle={2020 IEEE International Conference on Software Maintenance and Evolution (ICSME)}, 
  title={Effects of Adopting Code Review Bots on Pull Requests to OSS Projects}, 
  year={2020},
  volume={},
  number={},
  pages={1-11},
  doi={10.1109/ICSME46990.2020.00011},
  url={https://doi.org/10.1109/ICSME46990.2020.00011}}

@INPROCEEDINGS{antipatterninreview2021,
   author={Chouchen, Moataz and Ouni, Ali and Kula, Raula Gaikovina and Wang, Dong and Thongtanunam, Patanamon and Mkaouer, Mohamed Wiem and Matsumoto, Kenichi},
  booktitle={2021 IEEE International Conference on Software Analysis, Evolution and Reengineering (SANER)}, 
  title={Anti-patterns in Modern Code Review: Symptoms and Prevalence}, 
  year={2021},
  volume={},
  number={},
  pages={531-535},
  doi={10.1109/SANER50967.2021.00060},
  url={https://doi.org/10.1109/SANER50967.2021.00060}}

@misc{hassan2025agenticSE,
      title={Agentic Software Engineering: Foundational Pillars and a Research Roadmap}, 
      author={Ahmed E. Hassan and Hao Li and Dayi Lin and Bram Adams and Tse-Hsun Chen and Yutaro Kashiwa and Dong Qiu},
      year={2026},
      eprint={2509.06216},
      archivePrefix={arXiv},
      primaryClass={cs.SE},
      url={https://arxiv.org/abs/2509.06216}, 
}

@ARTICLE{silhoutescore,
   author={Kanwal, Moona and Khan, Muzammil Ahmad and Ismat, Najma and Khan, Najeed A. and Khan, Aftab A.},
  journal={IEEE Access}, 
  title={Machine Learning Approach to Classification of Online Users by Exploiting Information Seeking Behavior}, 
  year={2024},
  volume={12},
  number={},
  pages={53234-53249},
  doi={10.1109/ACCESS.2024.3383444},
  url={https://doi.org/10.1109/ACCESS.2024.3383444}
}

@article{o2007caution,
  author={O'brien, Robert M.},
title={A Caution Regarding Rules of Thumb for Variance Inflation Factors},
journal={Quality {\&} Quantity},
year={2007},
month={Oct},
day={01},
volume={41},
number={5},
pages={673-690},
issn={1573-7845},
doi={10.1007/s11135-006-9018-6},
url={https://doi.org/10.1007/s11135-006-9018-6}
}

@ARTICLE{justintime2013TSE,
  author={Kamei, Yasutaka and Shihab, Emad and Adams, Bram and Hassan, Ahmed E. and Mockus, Audris and Sinha, Anand and Ubayashi, Naoyasu},
  journal={IEEE Transactions on Software Engineering}, 
  title={A large-scale empirical study of just-in-time quality assurance}, 
  year={2013},
  volume={39},
  number={6},
  pages={757-773},
  doi={10.1109/TSE.2012.70},
  url={https://www.doi.org/10.1109/TSE.2012.70}}

@misc{llamapreview,
  author       = {{JetXu-LLM}},
  title        = {{LlamaPReview}: Evidence-Based, Low-Noise {AI} Code Reviewer},
  howpublished = {\url{https://github.com/marketplace/llamapreview}},
  note         = {GitHub Marketplace app page. Accessed: 2026-06-23}
}

@misc{repo_azurecli,
  title        = {{Azure Command-Line Interface}},
  howpublished = {\url{https://github.com/Azure/azure-cli}},
  note         = {Accessed: 2026-06-30}
}

@misc{repo_flutter,
  title        = {{Flutter Packages, a collection of useful packages maintained by the Flutter team}},
  howpublished = {\url{https://github.com/flutter/packages}},
  note         = {Accessed: 2026-06-30}
}

@article{senin2008dynamic,
  title = {A global averaging method for dynamic time warping, with applications to clustering},
journal = {Pattern Recognition},
volume = {44},
number = {3},
pages = {678-693},
year = {2011},
issn = {0031-3203},
doi = {https://doi.org/10.1016/j.patcog.2010.09.013},
url = {https://doi.org/10.1016/j.patcog.2010.09.013},
author = {François Petitjean and Alain Ketterlin and Pierre Gançarski}
}

@inproceedings{softdtw2017,
author = {Cuturi, Marco and Blondel, Mathieu},
title = {Soft-DTW: a differentiable loss function for time-series},
year = {2017},
publisher = {JMLR.org},
booktitle = {Proceedings of the 34th International Conference on Machine Learning - Volume 70},
pages = {894–903},
numpages = {10},
location = {Sydney, NSW, Australia},
series = {ICML'17},
url={https://dl.acm.org/doi/10.5555/3305381.3305474}
}

@misc{repo_rtthread,
  title        = {{RT-Thread, an open source IoT Real-Time Operating System (RTOS)}},
  howpublished = {\url{https://github.com/RT-Thread/rt-thread}},
  note         = {Accessed: 2026-06-30}
}

@misc{repo_openlibrary,
  title        = {{Open Library, an open, editable library catalog, building towards a web page for every book ever published.}},
  howpublished = {\url{https://github.com/internetarchive/openlibrary}},
  note         = {Accessed: 2026-06-30}
}

@misc{zhong2025developerllm,
      title={Developer-LLM Conversations: An Empirical Study of Interactions and Generated Code Quality}, 
      author={Suzhen Zhong and Ying Zou and Bram Adams},
      year={2025},
      eprint={2509.10402},
      archivePrefix={arXiv},
      primaryClass={cs.SE},
      url={https://arxiv.org/abs/2509.10402}, 
}

\end{document}